\documentclass[prb,twocolumn,showpacs,superscriptaddress,floatfix]{revtex4}
\usepackage{graphicx,amsfonts,amssymb,amsmath,hyperref,color}
\newif\ifhyper
\hypertrue
\ifhyper
\hypersetup{
  citecolor = {green},
  colorlinks = {true}, 
  urlcolor = {blue} 
}
\fi

\newcommand{\bn}{{\boldsymbol{n}}}

\definecolor{grau_hell}{gray}{0.9}
\definecolor{grau}{gray}{0.85}

\newcommand{\h}{H_{\!I}}
\newcommand{\hh}{H_{\!I\!I}}
\newcommand{\hhh}{H_{\!I\!I\!I}}

\newcommand{\gs}[2]{\mathcal{E}_{\mathrm{#1}}^{\mathrm{#2}}}

\begin{document}

\title{Kitaev model and dimer coverings on the honeycomb lattice}

\author{Michael Kamfor}
\email{kamfor@fkt.physik.tu-dortmund.de}
\affiliation{Lehrstuhl f\"ur Theoretische Physik I, TU Dortmund, Otto-Hahn-Stra\ss e 4, D-44221
  Dortmund, Germany}

\author{S\'ebastien Dusuel}
\email{sdusuel@gmail.com}
\affiliation{Lyc\'ee Saint-Louis, 44 Boulevard Saint-Michel, 75006 Paris, France}

\author{Julien Vidal}
\email{vidal@lptmc.jussieu.fr}
\affiliation{Laboratoire de Physique Th\'eorique de la Mati\`ere Condens\'ee,
CNRS UMR 7600, Universit\'e Pierre et Marie Curie, 4 Place Jussieu, 75252
Paris Cedex 05, France}

\author{Kai Phillip Schmidt}
\email{schmidt@fkt.physik.tu-dortmund.de}
\affiliation{Lehrstuhl f\"ur Theoretische Physik I, TU Dortmund, Otto-Hahn-Stra\ss e 4, D-44221 Dortmund, Germany}

\begin{abstract} 

We consider an extension of the Kitaev honeycomb model based on arbitrary dimer coverings satisfying matching rules. We focus on three different dimer coverings having the smallest unit cells for which we calculate the ground-state phase diagram. We also study one- and two-vortex properties for these coverings in the Abelian phases and show that vortex-vortex interactions can be attractive or repulsive. These qualitative differences are confirmed analytically by high-order perturbative expansions around the isolated-dimer limit. Similarities and differences with the original Kitaev honeycomb model are discussed.   

\end{abstract}

\pacs{75.10.-b, 75.10.Jm, 03.65.Vf, 05.30.Pr}

\maketitle

%
%
\section{Introduction}
\label{sec:intro}
%
%
Topologically ordered phases are fascinating states of quantum matter characterized by a ground-state degeneracy which depends on the genus of the considered topology \cite{Wen03,Wen04}. They are also known to exhibit exotic elementary excitations, dubbed anyons, which have nontrivial braiding statistics \cite{Leinaas77,Wilczek82_1,Wilczek82_3}. 
Recently, the interest for the study of  topologically ordered phases has been triggered by the possibility to use anyons as an essential ingredient either to perform quantum computation or to provide a quantum memory \cite{Preskill_HP}.
To understand these properties, much focus has been drawn on two-dimensional quantum spin systems especially since Kitaev has proposed two exactly solvable models, the toric code \cite{Kitaev03} and the honeycomb model \cite{Kitaev06},  in which such anyons are present. 

In the present work, we provide an extension of the Kitaev honeycomb model for which many experimental realizations in optical lattices \cite{Duan03,Micheli06,Jiang08}, Josephson junctions arrays \cite{You10}, or solid state systems with strong spin-orbit coupling \cite{Jackeli09,Chaloupka10} have been proposed. 
Up to now, many variations of this model have already been investigated, but almost all of them rely on changing the geometry of the underlying lattice \cite{Yao07,Yang07,Dusuel08_2,Mandal09,Nussinov09,Baskaran09,Nash09}. 
Here, we choose an alternative route by keeping the honeycomb geometry but by changing the coupling configuration. Doing so, we 
propose a set of exactly solvable models which, as we will see, can give rise to surprising properties as compared to the original one \cite{Kitaev06}. 

This paper is organized as follows. In Sec.~\ref{sec:themodel} we recall the basic properties of the Kitaev honeycomb model and we explain how to extend it by considering different dimer coverings.  In Sec.~\ref{sec:phasediag}, we focus on three simple coupling patterns for which we compute the ground-state (vortex-free) phase diagrams using Majorana fermionization and analyze the influence of a three-spin interaction term. The following Sec.~\ref{sec:vortexprop} is dedicated  to sparse vortex configurations. There, we use a perturbative treatment to show that the nature (attractive or repulsive) of vortex-vortex interaction is very sensitive to the dimer configuration.

%
%
\section{Kitaev model and dimer coverings}
\label{sec:themodel}
\begin{figure}[h]
\includegraphics[width=0.5\columnwidth]{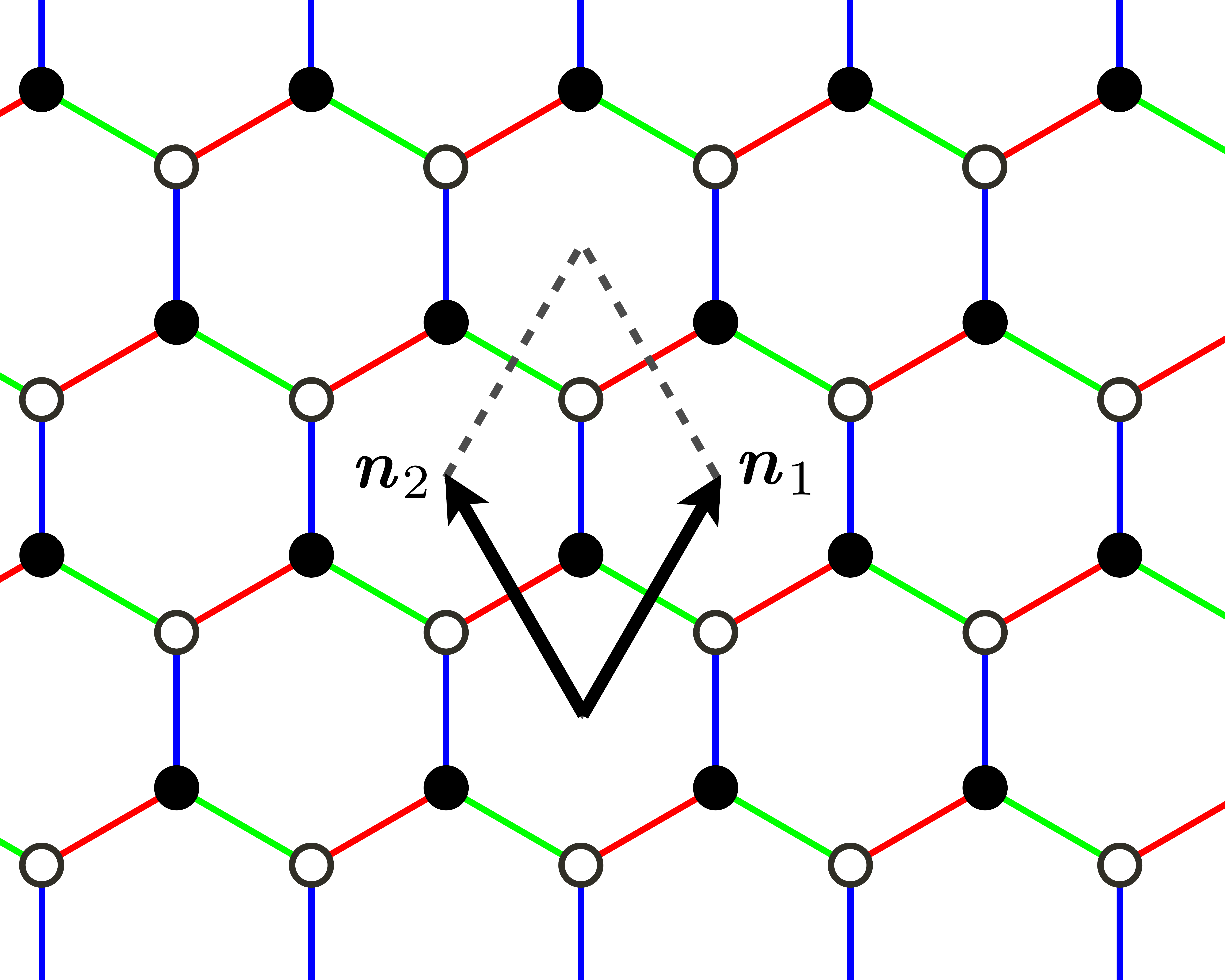} 
\caption{Dimer covering $I$ initially considered by Kitaev \cite{Kitaev06} with two sites per unit cell.  Vectors $\bn_1$ and $\bn_2$ define the unit cell. Red, green, and blue links represent $x, y$, and $z$ links respectively.}
\label{fig:lattice1}
\end{figure}

%
%
The Kitaev honeycomb model describes a set of spins 1/2 located at the vertices of a hexagonal lattice which interact via the following Hamiltonian
%
%
\begin{equation}
  \label{eq:ham}
  H=-\sum_{\alpha=x,y,z}\sum_{\alpha-\mathrm{links}}
  J_\alpha\,\sigma_i^\alpha\sigma_j^\alpha,
\end{equation} 
%
%
where $\sigma_i^\alpha$ are the usual Pauli matrices at site $i$.  In the original model \cite{Kitaev06}, Kitaev considered the case where the type of links ($\alpha=x,y$, or $z$) depends only on its orientation as depicted in Fig.~\ref{fig:lattice1}.

%
%
%
%

So far, extensions of this model were based on a different lattice geometry \cite{Yao07,Yang07,Dusuel08_2,Mandal09,Baskaran09} or a different underlying algebra \cite{Nussinov09,Yao09}. 
Sticking to a spin-1/2 model,  the exact solvability is preserved only if the following two constraints are satisfied: $(i)$ each site of the system must be trivalent and $(ii)$ the three links connected to
a given site have to be of different types ($x$, $y$, and $z$).
Thus, the Hamiltonian (\ref{eq:ham}) can be defined and solved on any trivalent graph.

\begin{figure}[htb]
\includegraphics[width=\columnwidth]{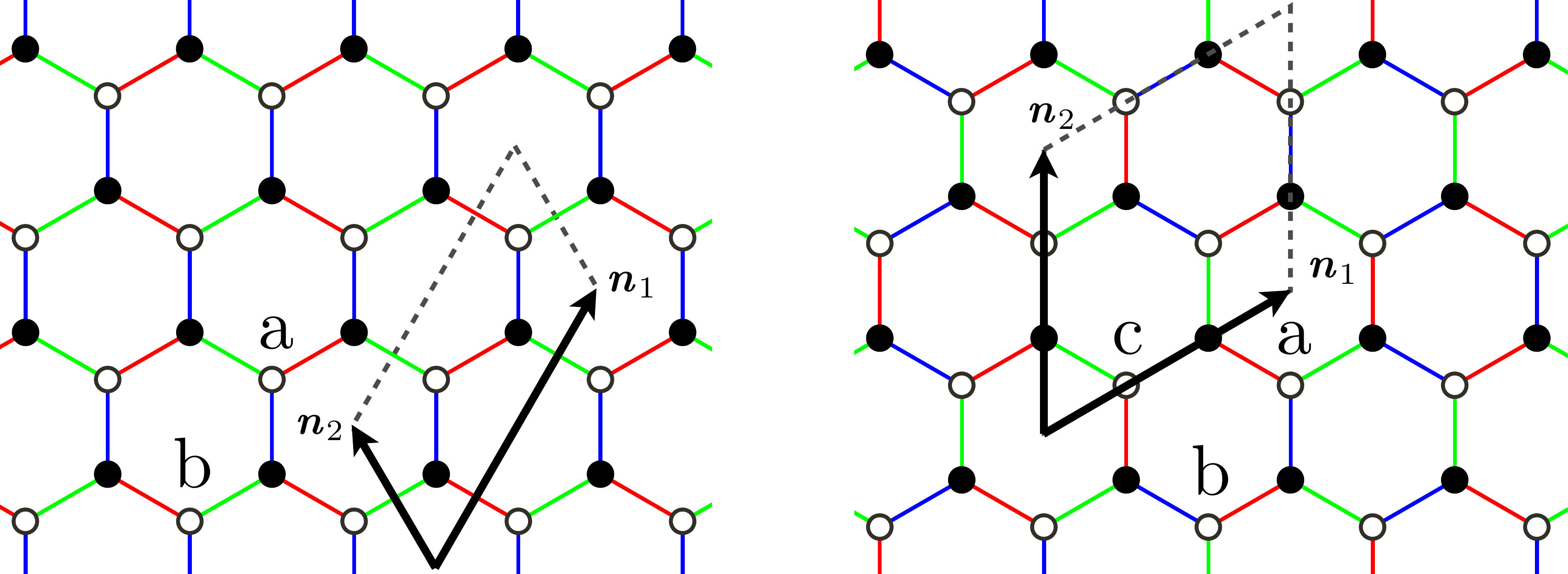}
\caption{Left: dimer covering ${I\!I}$ with four sites per unit cell. Right: dimer covering ${I\!I\!I}$ with six sites per unit cell. Conventions are the same as in Fig.~\ref{fig:lattice1} and labels (a), (b), and (c) refer to the different kinds of plaquettes whose corresponding operators are given in Eqs.~(\ref{eq:Wp_II}-\ref{eq:Wp_III}).}
\label{fig:lattice2}
\end{figure}

Here, by contrast, we study the Kitaev model (\ref{eq:ham}) in the original honeycomb lattice but for different link distributions or dimer coverings (we shall use the expressions link distributions or dimer coverings as synonyms, so dimer coverings do not have the same meaning here as they have in the context of quantum dimer models). More precisely, we focus on the three simple dimer coverings shown in Figs.~\ref{fig:lattice1} and \ref{fig:lattice2}. These configurations are among the simplest periodic decorations one may build on such a lattice. Indeed, the honeycomb lattice has two sites per unit cell so that any periodic decoration will have $2 n$ sites per unit cell. Configurations shown in Figs.~\ref{fig:lattice1}-\ref{fig:lattice2} correspond to $n=1,2$, and 3 respectively, and the corresponding models will be consequently referred to as $I$, ${I\!I}$, and ${I\!I\!I}$. In this terminology, the original Kitaev model introduced in Ref.~\onlinecite{Kitaev06}  is model $I$.
%
%
%
%

As shown below, the phase diagram strongly depends on the dimer covering and displays very different properties when switching on a three-spin interaction term already introduced by Kitaev~\cite{Kitaev06} to break the time-reversal symmetry explicitely. Such a term is built in the following way: if two dimers $(i,j)$ of type $\alpha$ and $(j,k)$ of type $\beta$ share a common site $j$, then the three-spin term is defined by $\sigma_i^\alpha\sigma_j^\gamma \sigma_k^\beta$ where $\gamma$ is the direction orthogonal to $\alpha$ and $\beta$.

The Hamiltonian considered in the present work can thus be written as
%
%
\begin{equation}
  \label{eq:hamK}
  H=-\sum_{\alpha=x,y,z}\sum_{(i,j)_\alpha}
  J_\alpha\,\sigma_i^\alpha\sigma_j^\alpha -K\!\!\! \sum_{(i,j)_\alpha,(j,k)_\beta} \sigma_i^\alpha \sigma_j^\gamma \sigma_k^\beta.
\end{equation} 
%
%
Here, we only  consider positive couplings $J_\alpha$ and $K$.
A crucial feature of Hamiltonian~(\ref{eq:hamK}) is that it commutes with plaquette operators defined, on each plaquette $p$, as $W_p=\prod_i\sigma_i^{\mathrm{out}(i)}$ where $i$ runs over the set of six
spins around the plaquette $p$, and where the notation $\mathrm{out}(i)$ means
the ``outgoing'' direction at site $i$, with respect to the plaquette's contour \cite{Vidal08_2}. Since $W_{p}^2 = \mathbb{I}$,  plaquette operators $W_{p}$ have eigenvalues $w_p = \pm 1$. Consequently, there is one conserved $\mathbb{Z}_2$ degree of freedom per plaquette. Following Kitaev \cite{Kitaev06}, we will say that there is a vortex on a plaquette if  $w_p = -1$ and no vortex if $w_p = +1$. Since $[H,W_{p}]=0$, one can block-diagonalize $H$ in each vortex sector given by a configuration of the $w_p$'s. 

The three dimer coverings considered here differ by the number and the type of plaquette operators $W_p$. More precisely, for covering $I$ there is only one type of plaquette operator
\begin{equation}
 \label{eq:Wp_I}
  W_p^{I}= \sigma_1^{z}\sigma_2^{x}\sigma_3^{y}\sigma_4^{z}\sigma_5^{x}\sigma_6^{y} ,
\end{equation}
where $1$ to $6$ are the six spins numbered clockwise around any plaquette $p$, spin number $1$ being at the bottom of the plaquette. Covering $I\!I$ has two types (a and b) of plaquette operators (see Fig.~\ref{fig:lattice2})
\begin{eqnarray}
 \label{eq:Wp_II}
  W_{p_\mathrm{a}}^{I\!I,\mathrm{a}} &=& \sigma_1^{z}\sigma_2^{x}\sigma_3^{x}\sigma_4^{z}\sigma_5^{y}\sigma_6^{y},\\
  W_{p_\mathrm{b}}^{I\!I,\mathrm{b}} &=& \sigma_1^{z}\sigma_2^{y}\sigma_3^{y}\sigma_4^{z}\sigma_5^{x}\sigma_6^{x}.
\end{eqnarray}
Covering $I\!I\!I$ has three types (a, b and c) of plaquette operators (see Fig.~\ref{fig:lattice2})
\begin{eqnarray}
  W_{p_\mathrm{a}}^{I\!I\!I,\mathrm{a}} &=& \sigma_1^{z}\sigma_2^{z}\sigma_3^{z}\sigma_4^{z}\sigma_5^{z}\sigma_6^{z}, \\
  W_{p_\mathrm{b}}^{I\!I\!I,\mathrm{b}} &=& \sigma_1^{y}\sigma_2^{y}\sigma_3^{y}\sigma_4^{y}\sigma_5^{y}\sigma_6^{y}, \\
 \label{eq:Wp_III}
   W_{p_\mathrm{c}}^{I\!I\!I,\mathrm{c}} &=& \sigma_1^{x}\sigma_2^{x}\sigma_3^{x}\sigma_4^{x}\sigma_5^{x}\sigma_6^{x} .
\end{eqnarray}
One can readily see that the structure of these operators is similar for coverings $I$ and $I\!I$ since they contain two Pauli matrices of each kind ($x, y$ and $z$) whereas, for covering $I\!I\!I$, each plaquette operator is made of identical Pauli matrices.
%
%
\section{Phase Diagrams}
\label{sec:phasediag}
%
%
For the three dimer coverings introduced in Sec.~\ref{sec:themodel}, we shall now determine the ground-state phase diagram. As shown by Lieb in a different context \cite{Lieb94}, the ground state, at least for $K=0$, lies in the vortex-free sector ($w_p$=$+1$ for all $p$) on which we focus thereafter.

\subsection{Majorana fermionization}
\label{ssec:technique}

Undoubtedly, one of the most remarkable properties of the Hamiltonian (\ref{eq:hamK}) is that, in each vortex sector,  it can be mapped onto a free Majorana fermion problem in a static $\mathbb{Z}_2$ gauge field. 
As detailed in Ref.~\onlinecite{Kitaev06}, $H$ can be written in terms of Majorana fermion operators $c_j$ at site $j$, and reads
%
%
\begin{equation}
\label{eq:trafohami}
H = \frac{\mathrm{i}}{4} \sum_{j,k} A_{jk} c_j c_k, 
\end{equation} 
%
%
where $A_{jk} = 2 J_{jk} u_{jk} + 2 K u_{ji}  u_{ki}$. In the latter expression, $J_{jk}$ is the coupling on the link $(j,k)$ and $u_{jk}=-u_{kj}=\pm 1$ is a gauge field which describes the vortex configuration through the relation $w_p=\prod_{(j,k) \in p} u_{jk}$ where $j$ belongs to the black sublattice and $k$ to the white one (see Figs.~\ref{fig:lattice1} and \ref{fig:lattice2}). In Eq.~(\ref{eq:trafohami}), the sum runs over all sites and $u_{ij}$ is only nonvanishing when $i$ and $j$ are nearest neighbors. 

Thus, for a given vortex configuration, one simply has to diagonalize the matrix $A$ to get the spectrum of the corresponding sector. Obviously, such a task is easy for periodic configurations but difficult otherwise. In particular, when one is interested in sparse vortex configurations, one needs to diagonalize large systems and perform some finite-size analysis (see Sec.~\ref{sec:vortexprop}).

To determine the phase diagram, we computed the fermionic gap for the three dimer coverings $I$, $I\!I$, and $I\!I\!I$. For symmetry reasons, all results below are displayed in the plane $J_x+J_y+J_z=1$ which is further parametrized in the following way 
\begin{eqnarray}
 x&=&1-J_x+J_y,\\
 y&=&1-J_x-J_y=J_z,
\end{eqnarray} 
with $x,y\in\{0,1\}$.

%
%
\subsection{Dimer covering $I$}
\label{sec:nu0HI}
%
%
The dimer covering shown in Fig.~\ref{fig:lattice1} is completely symmetric under the exchange of the three directions $x,y$, and $z$. Thus, for convenience and without loss of generality, let us assume $J_z \geqslant J_x,J_y$. The fermionic gap as a function of $K$ evolves as follows. \\

$\bullet$ For $K=0$,  one has \cite{Kitaev06}
%
%
\begin{equation}
\label{eq:gap10}
\Delta_I (K=0)=\mathrm{Max}\{2(J_z-J_x-J_y),0\},
\end{equation}
%
%
so that one has to distinguish between a gapped phase $A$ for $J_z>J_x+J_y$ and a gapless phase $B$ for 
$J_z<J_x+J_y$. \\

$\bullet$ In the small $K$ limit, a gap opens up in the $B$ phase whereas, in the $A$ phase,  the gap does not depend on $K$. For all $K$, the spectrum remains gapless on the line $J_z=J_x+J_y$ \cite{Lahtinen08}.  \\

\begin{figure}[t]
	\centering
  \includegraphics[width=0.4\columnwidth]{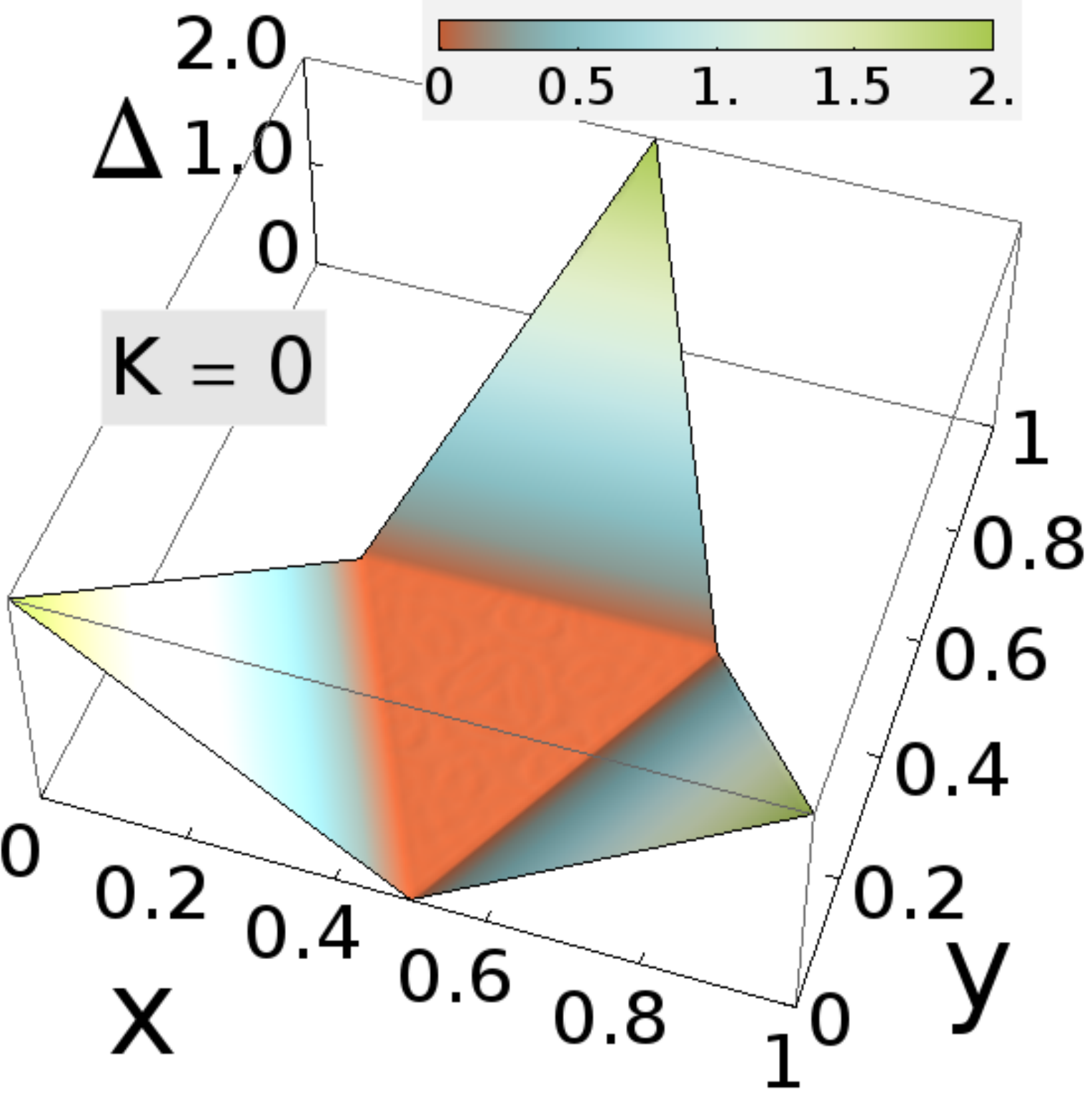}
  \includegraphics[width=0.4\columnwidth]{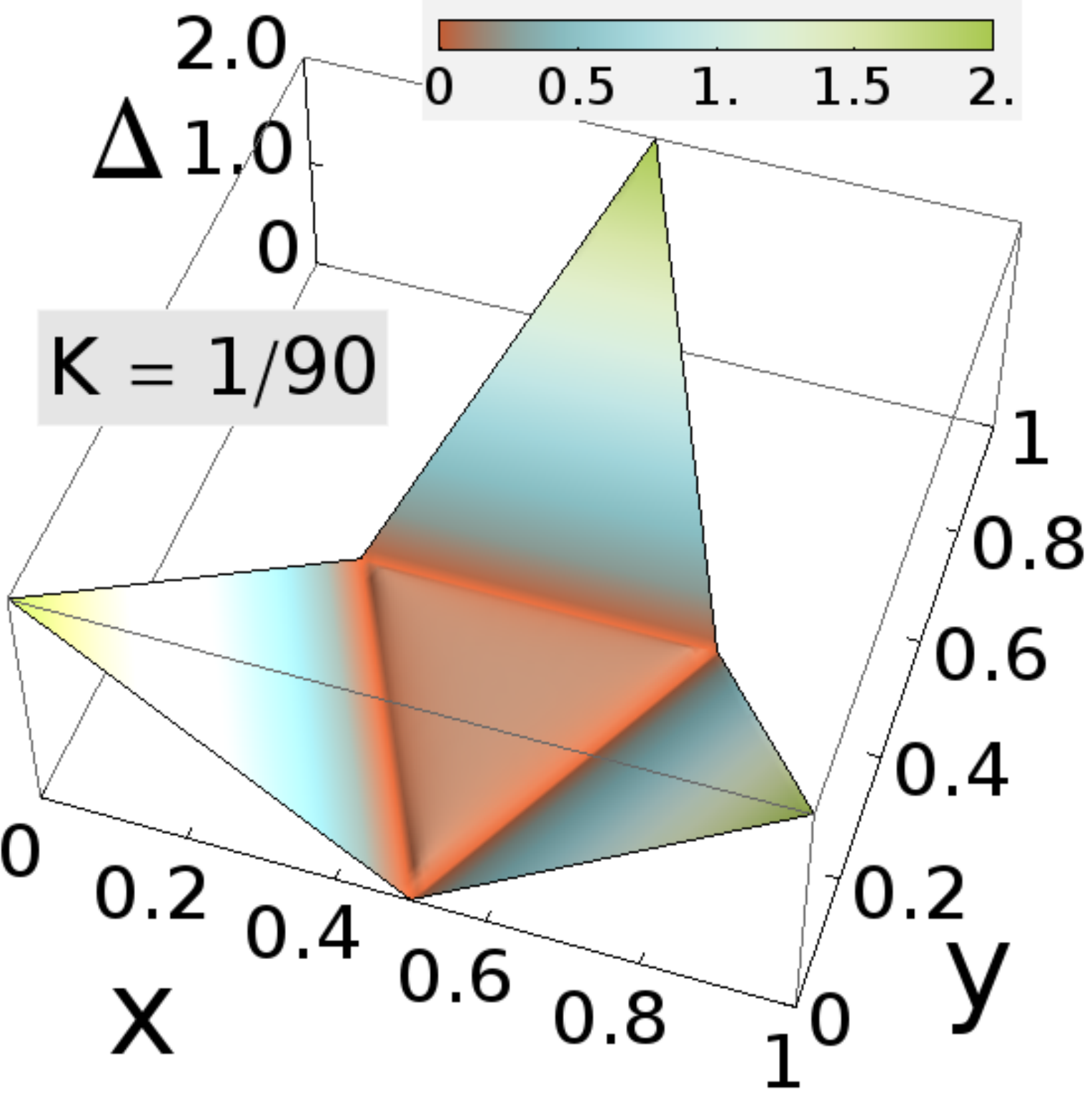}\\
  \includegraphics[width=0.4\columnwidth]{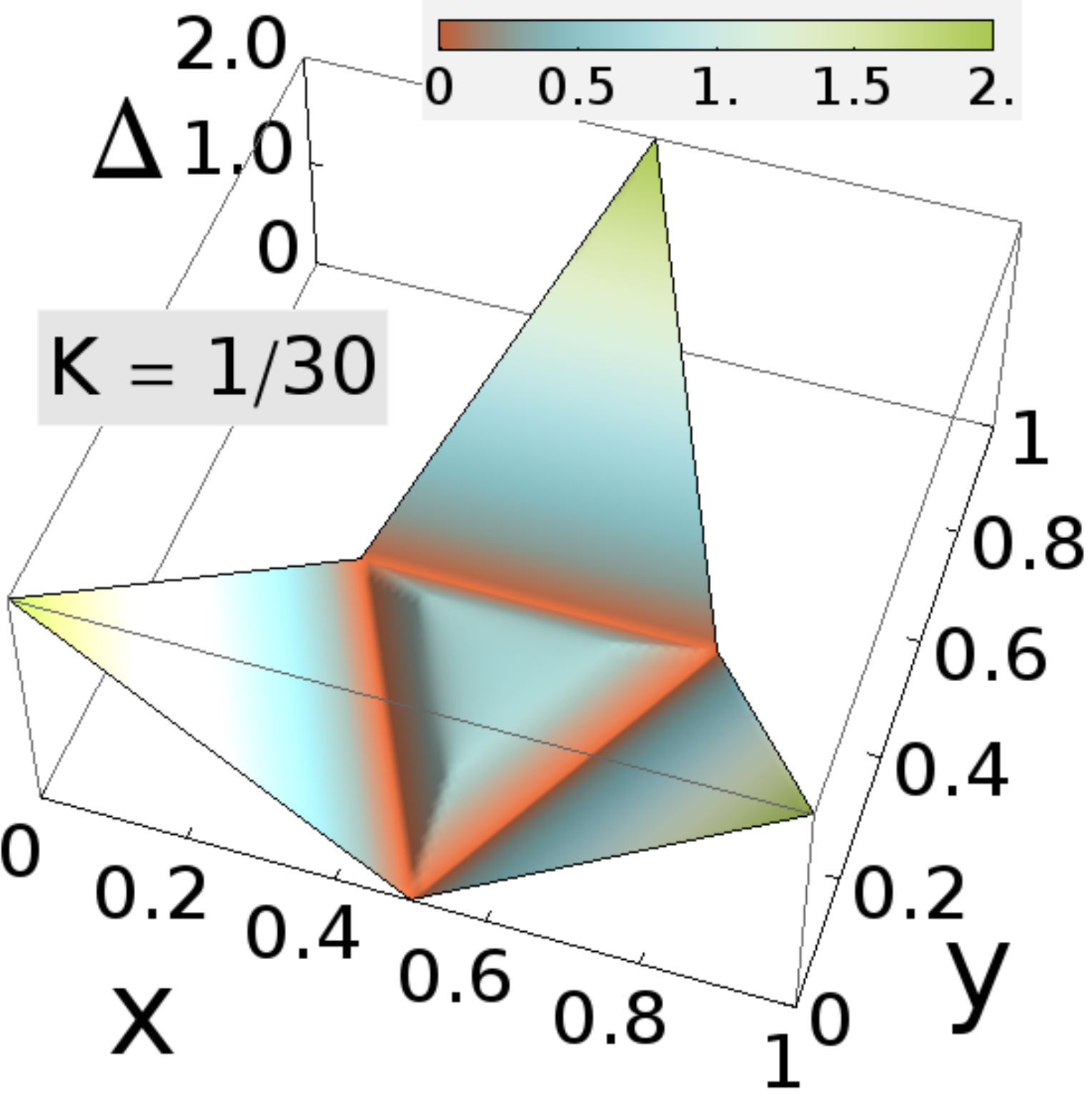}
  \includegraphics[width=0.4\columnwidth]{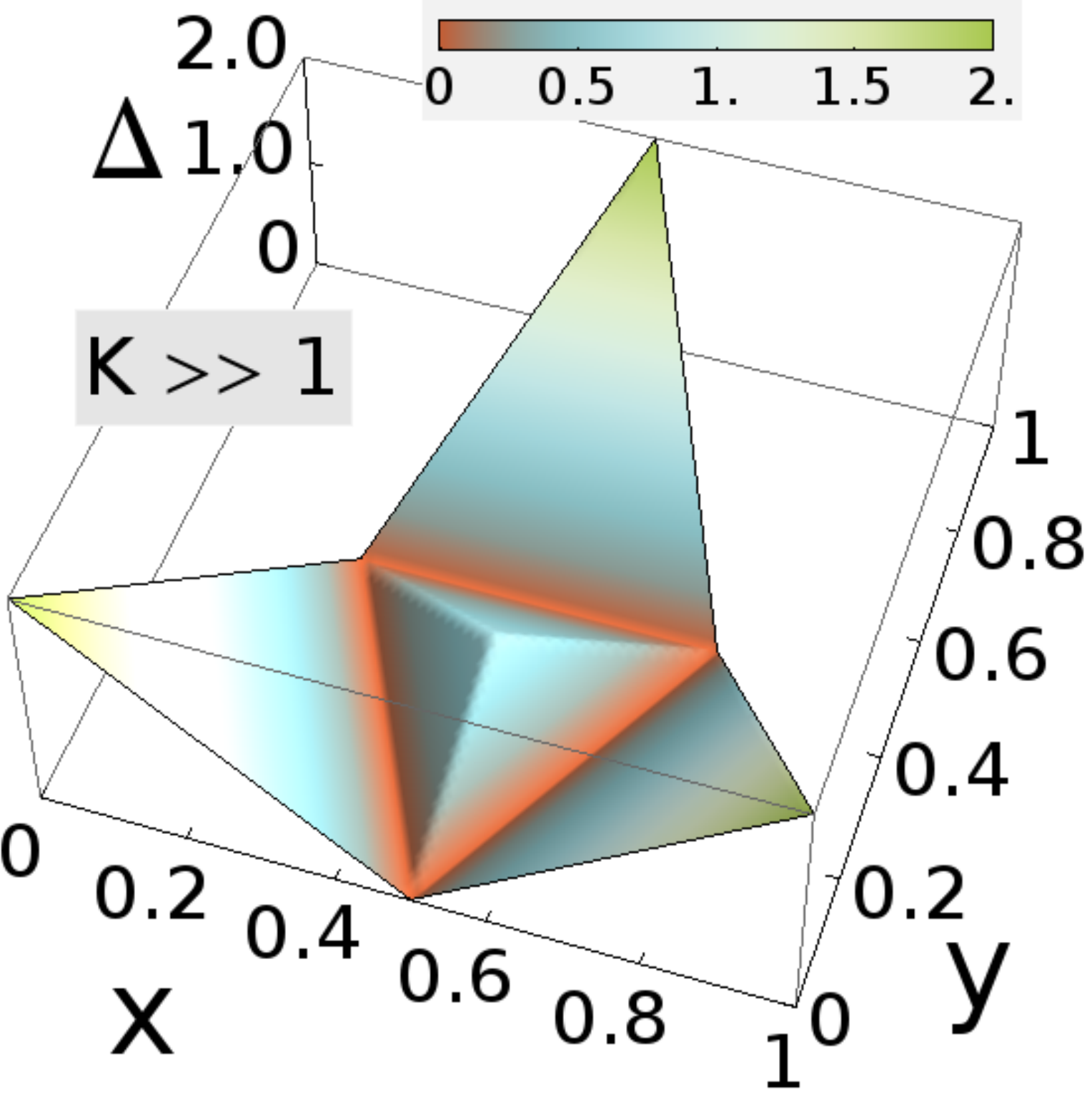}
\caption{Evolution of the fermionic gap in the vortex-free sector for covering $I$ as a function of $K$. Gapped regions at $K=0$ correspond to phases A and the gapless domain is identified with phase B. For finite $K$ the phase B becomes gapped and the gap saturates in the large $K$ limit.}
  \label{fig:square0}
\end{figure}

$\bullet$ In the large $K$ limit, the gap in the $B$ phase reaches the following asymptotic value 
%
%
\begin{equation}
\label{eq:gap10b}
\Delta_I (K=\infty)=2|J_z-J_x-J_y|. 
\end{equation}
%
%
The evolution  of the phase diagram when varying $K$ is given in Fig.~\ref{fig:square0}.

%
%
%
%
As shown by Kitaev, phase A is Abelian, whereas phase B contains non-Abelian excitations in the presence of a three-spin term \cite{Kitaev06}. In both cases, anyonic excitations are vortices living on plaquettes.

%
%
\subsection{Dimer covering $I\!I$}
\label{sec:nu0HII}
%
%
Contrary to covering $I$, covering $I\!I$ is not symmetric under the exchange of the three directions $x,y$, and $z$, but it is only symmetric under the exchange $x\leftrightarrow y$.  In this case, the evolution of the fermionic gap with $K$ is the following.\\

$\bullet$ For $K=0$, the phase diagram is exactly the same as for dimer covering $I$. Since model $I\!I$ is equivalent to model $I$ when $J_x=J_y=0$ or $J_x=J_y=J_z$, one can conclude that the statistics of the excitations in phases $A$ and $B$ will also correspond to the statistics for covering $I$. The same argument will hold for model $I\!I\!I$.\\

$\bullet$  For $K \neq 0$, the situation is more complex and one must distinguish between two regions. \\

- For  $J_z>J_x+J_y$, the gap is independent of $K$ and is still given by Eq.~(\ref{eq:gap10}).\\

- For  $J_z<J_x+J_y$, the gap has a nontrivial $K$-dependence. It vanishes on the lines 
%
%
\begin{equation}
J_z^2+16 K^2=(J_x-J_y)^2.
\end{equation}
%
%

\begin{figure}[t]
	\centering
  \includegraphics[width=0.4\columnwidth]{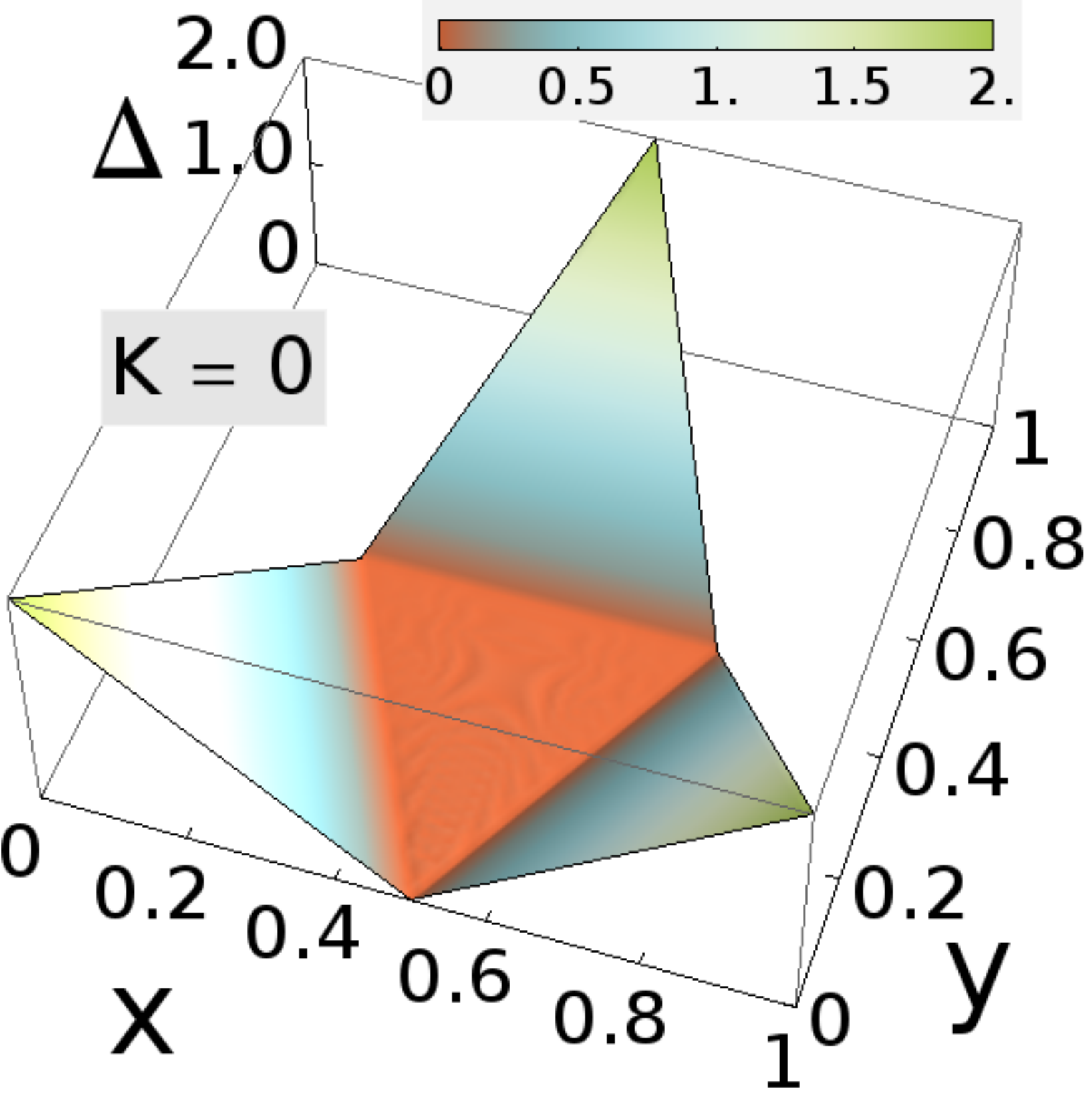}
  \includegraphics[width=0.4\columnwidth]{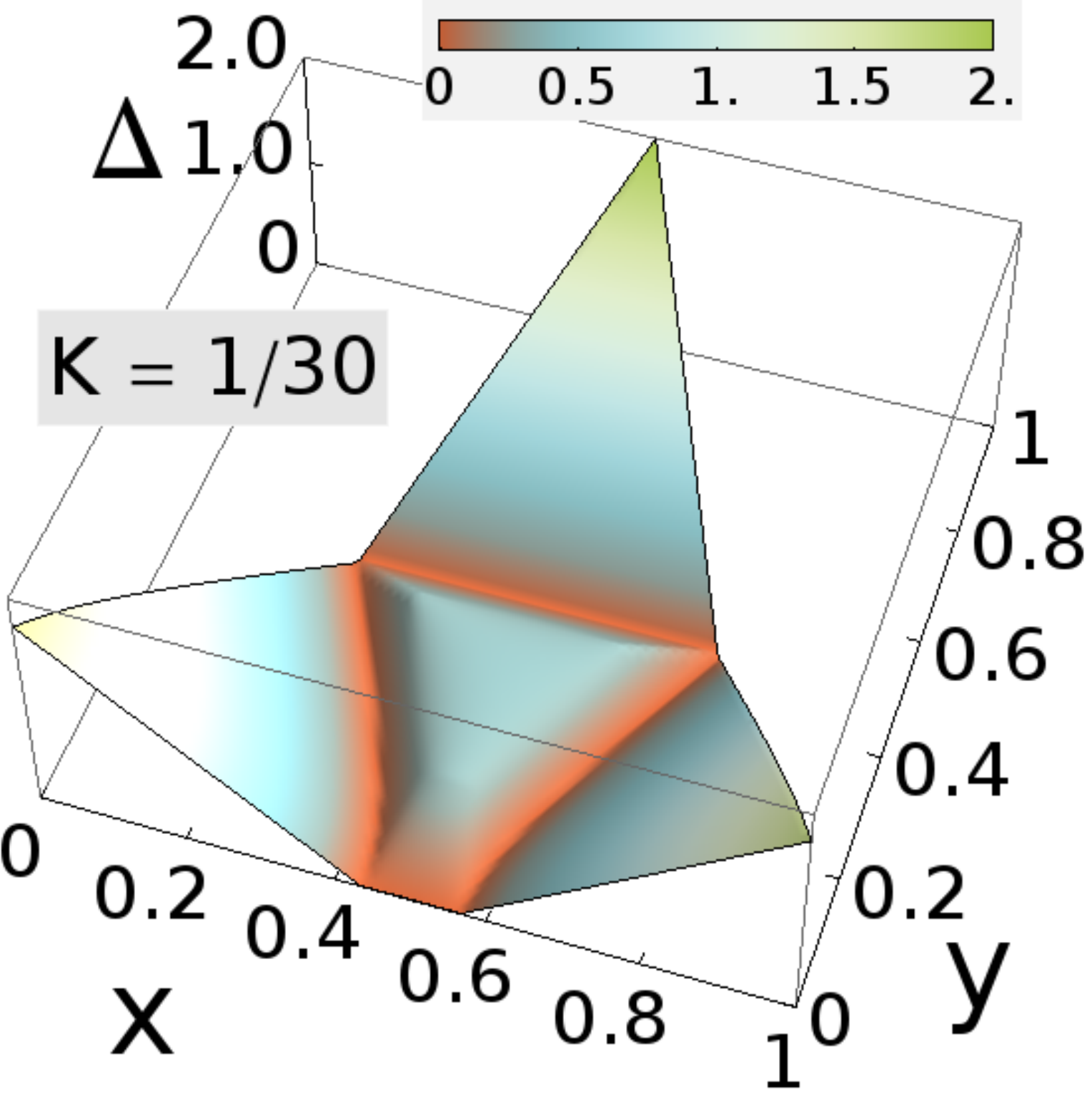}\\
  \includegraphics[width=0.4\columnwidth]{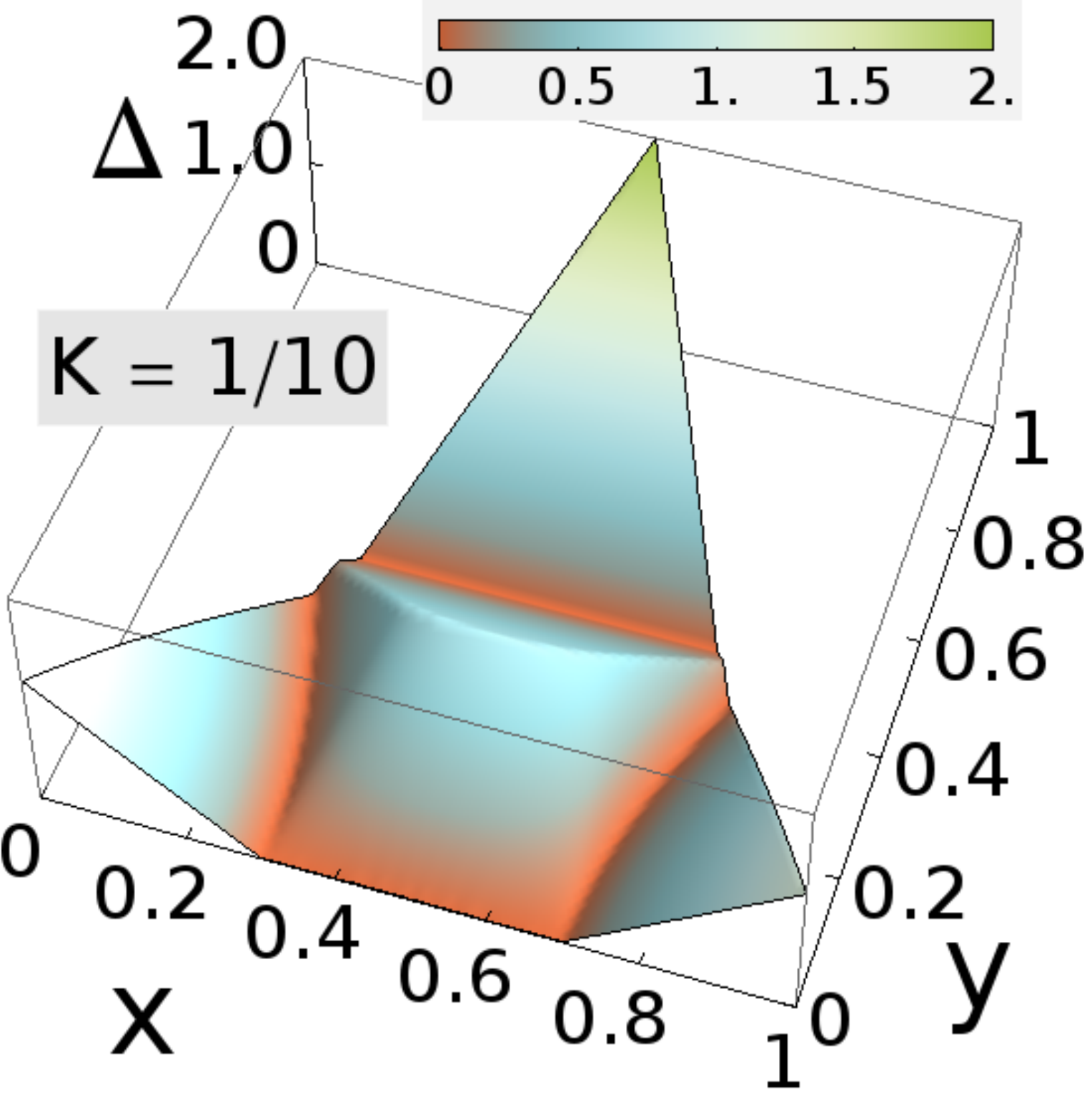}
  \includegraphics[width=0.4\columnwidth]{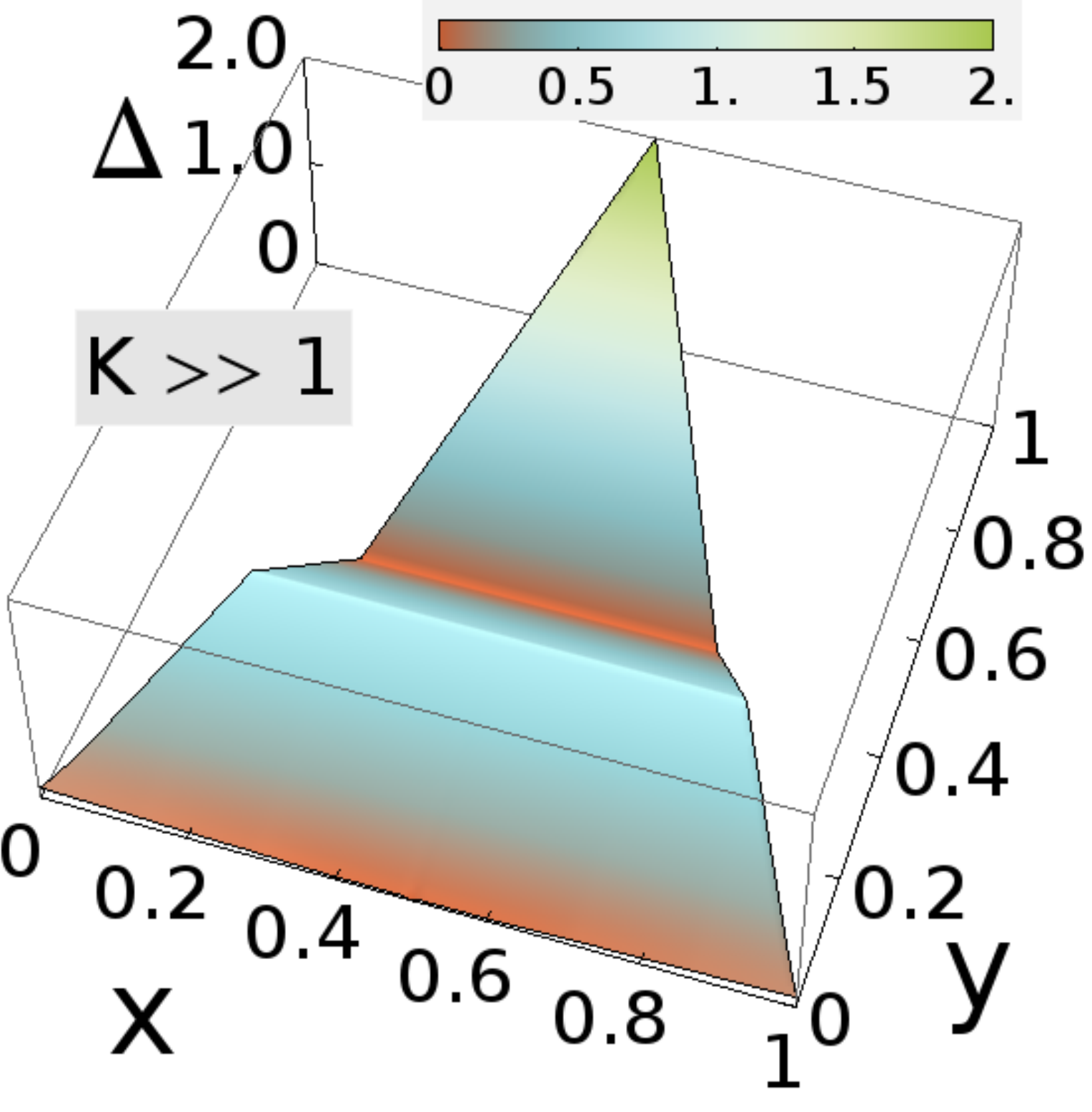}
\caption{Evolution of the fermionic gap in the vortex-free sector for covering $I\!I$ as a function of $K$. For $K=0$, the gap is the same as for covering $I$.}
  \label{fig:HII}
\end{figure}

$\bullet$ In the large $K$ limit, the gap for $J_z>J_x+J_y$ remains unchanged whereas the gap in the  $B$ phase is made up of two planes which intersects on the line $J_x+J_y=2/3$ (see Fig.~\ref{fig:HII}).

%
%
%
%

%
%
\subsection{Dimer covering $I\!I\!I$}
\label{sec:nu0HIII}
\begin{figure}[t]
	\centering
  \includegraphics[width=0.4\columnwidth]{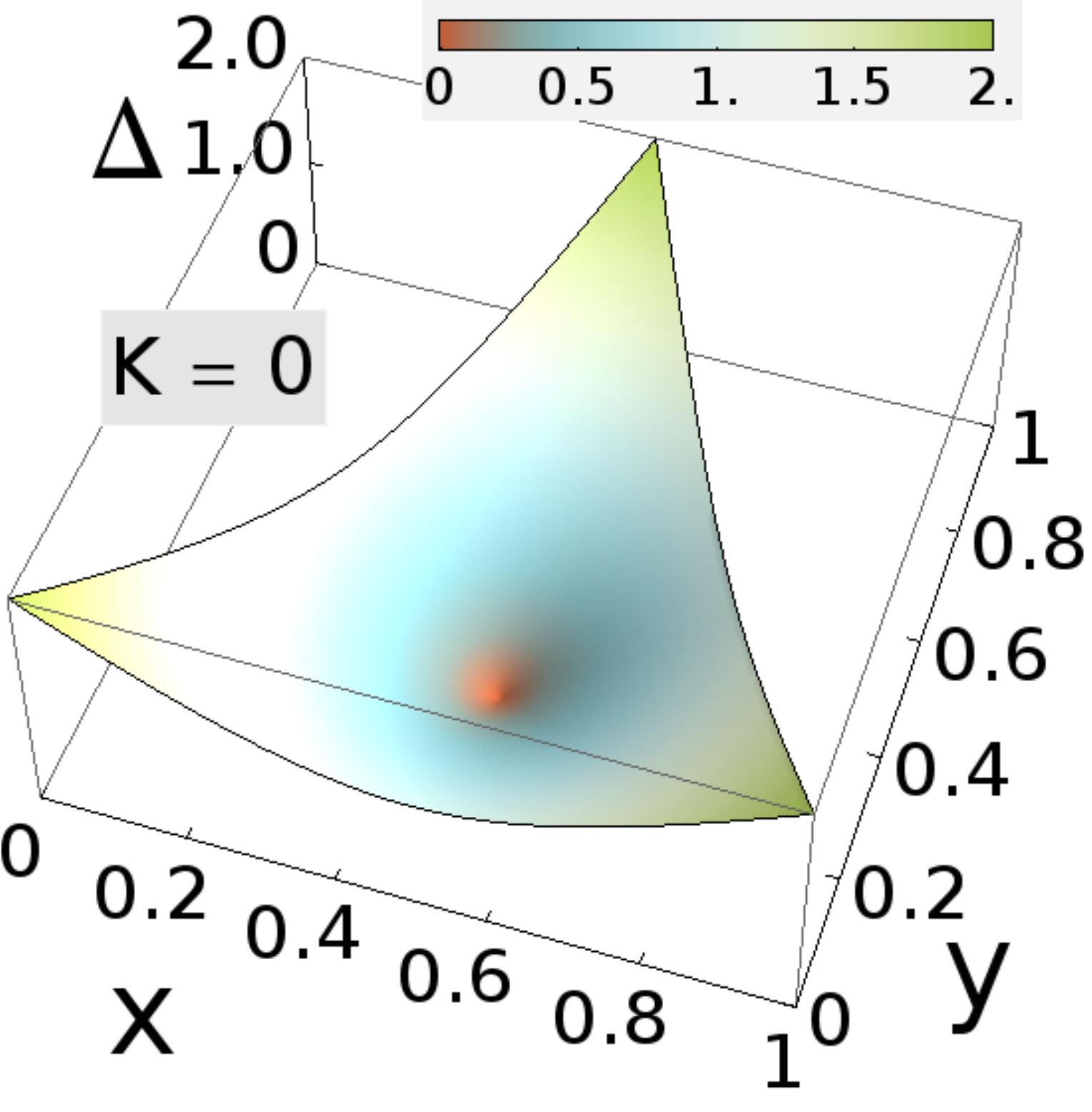}
  \includegraphics[width=0.4\columnwidth]{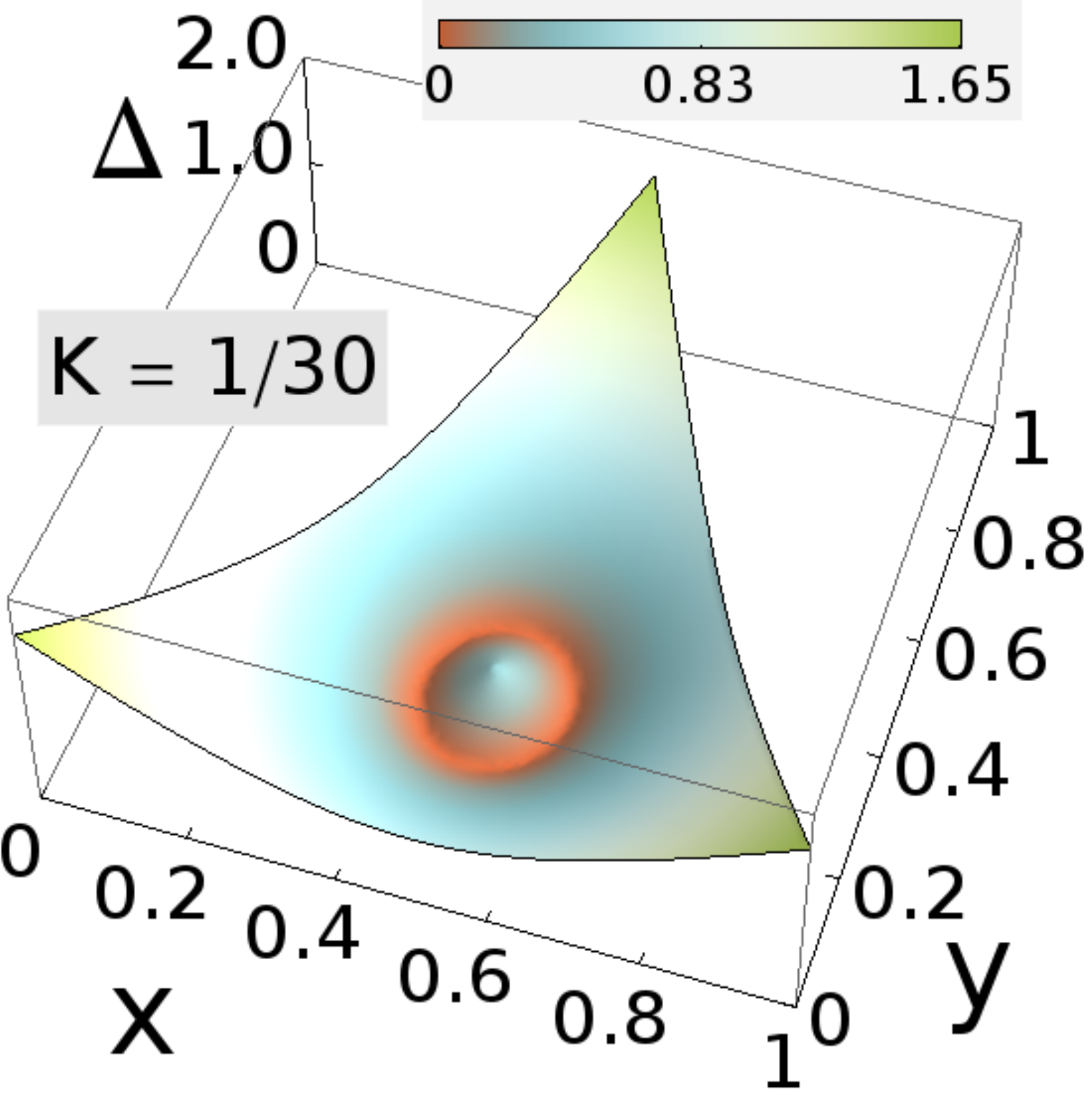}\\
  \includegraphics[width=0.4\columnwidth]{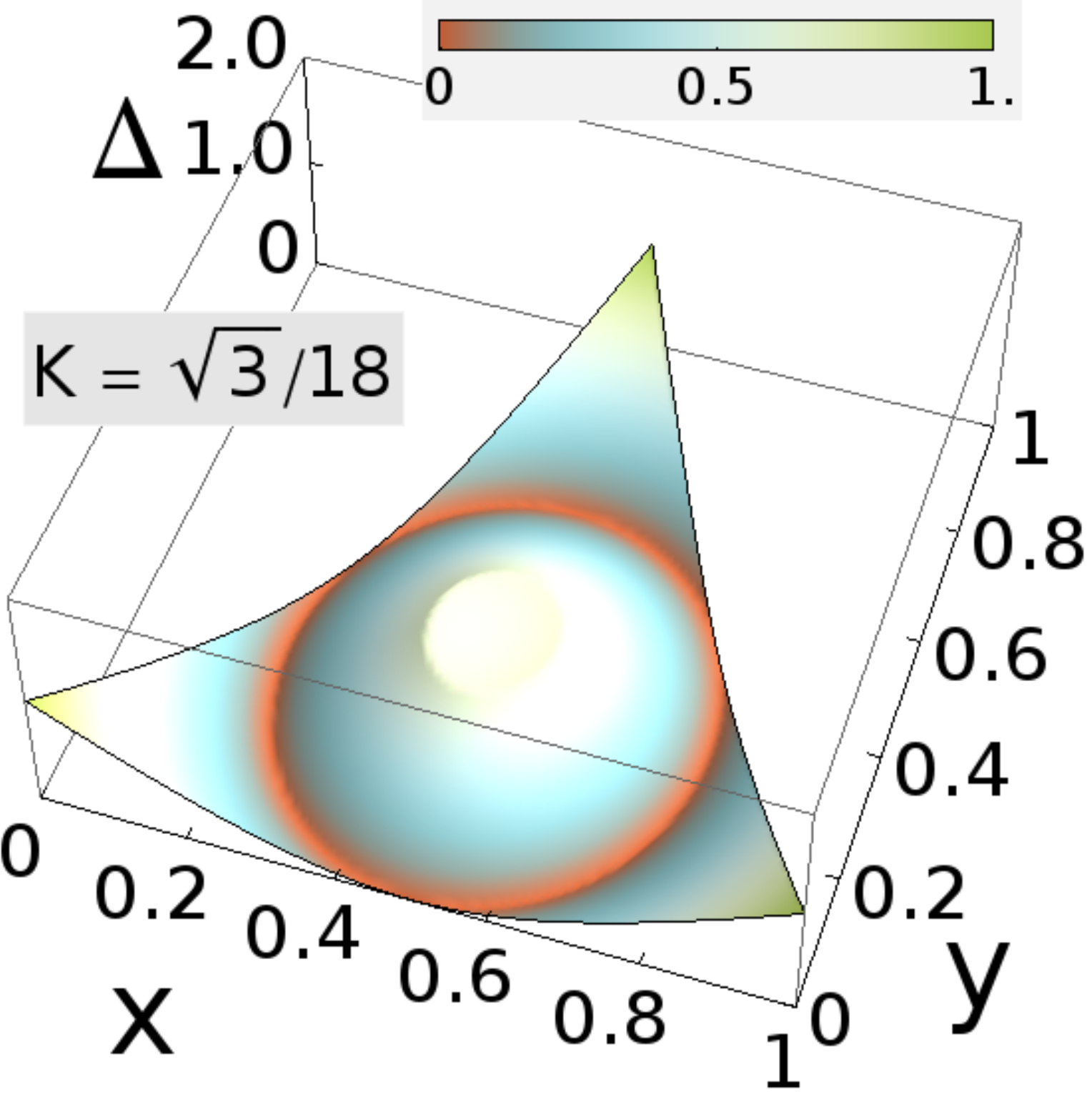}
  \includegraphics[width=0.4\columnwidth]{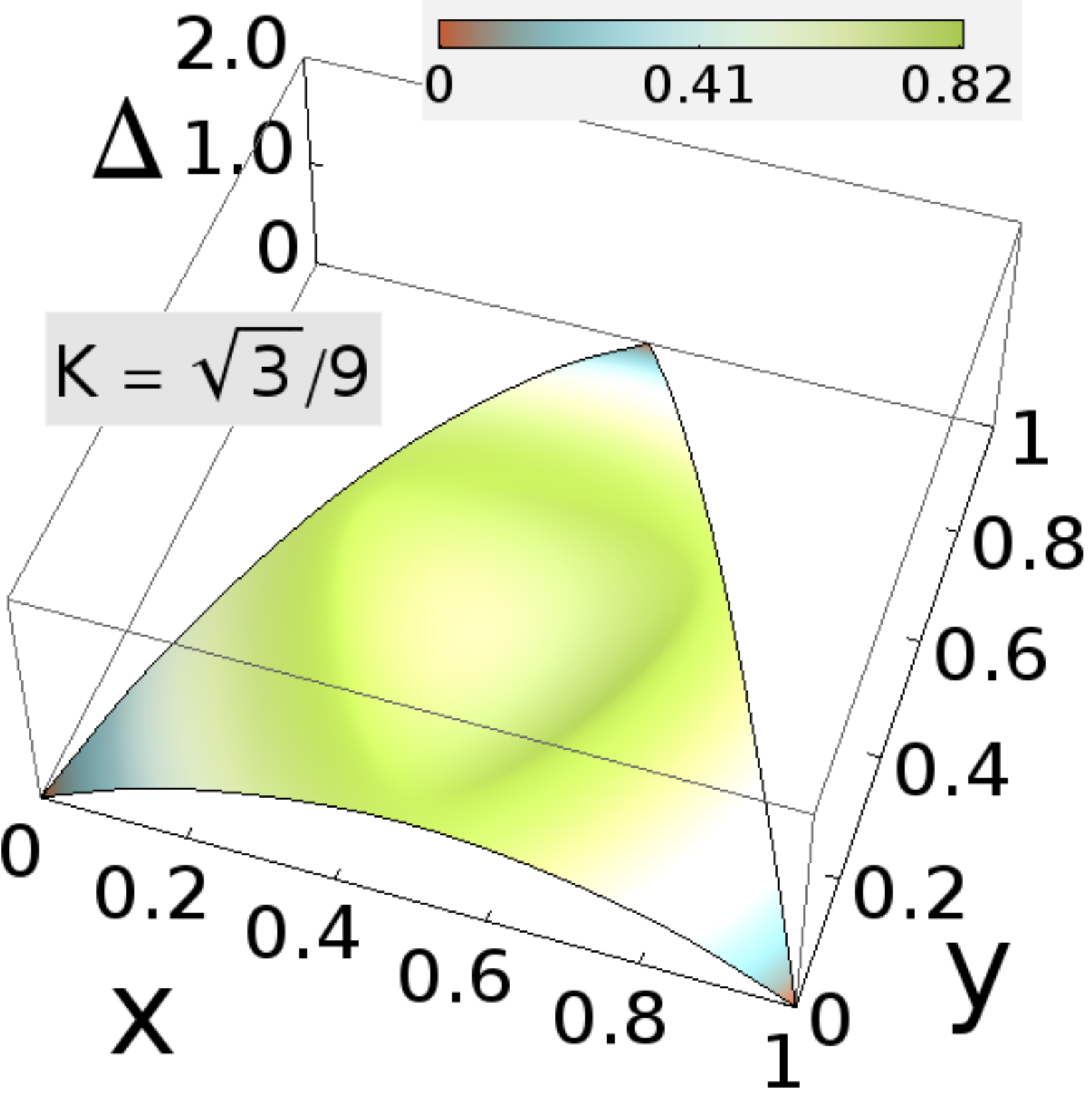}
\caption{Evolution of the fermionic gap in the vortex-free sector for covering $\!I\!I\!I$ as a function of $K$.  For $K=0$,  there is a single gapless point and the whole system is in an Abelian phase. By contrast, in the large $K$ limit, the system is in a non-Abelian phase and the gap is constant.}
  \label{fig:kagome0}
\end{figure}

%
%
This covering is completely symmetric under the exchange of the three directions $x,y$, and $z$. The fermionic gap as a function of $K$ is however very different from previous cases, and is illustrated in Fig.~\ref{fig:kagome0}.\\

$\bullet$ For $K=0$, the gap is given by
%
%
\begin{equation}
\Delta_{\!I\!I\!I} (K=0) = 2 \sqrt{J_x^2+J_y^2+J_z^2 -J_xJ_y -J_yJ_z-J_xJ_z}.
\end{equation}
%
%
As can be easily checked, it only vanishes at the isotropic point $J_x=J_y=J_z$ (as besides for any possible covering of the honeycomb lattice at $K=0$). In other words, for this covering there is no gapless phase but just a gapless point. \\

$\bullet$ For $K\neq0$, a gapped non-Abelian phase starts to develop around the isotropic point delimited by a gapless circle.  The gap in the Abelian region decreases when increasing $K$. \\
%
%
%
%

- For $K=\sqrt{3}/18$, the gapless circle reaches the middle of the triangle sides and isolates each Abelian gapped phase.  \\

- For $K=\sqrt{3}/9$, the gapless line finally reaches the triangle corners so that only the non-Abelian phase remains. \\

$\bullet$ In the large-$K$ limit, the gap becomes constant and equal to $\Delta_{\!I\!I\!I} (K=\infty)=2/3$ in the whole parameter range considered and the excitations are non-Abelian anyons.
%
%
\section{Vortex Properties}
\label{sec:vortexprop}
%
%

Let us now discuss the one- and two vortex sectors. Our main motivation for such a study is to unveil some qualitative differences of the vortex properties for the different coverings. Such differences are already present in the Abelian phase ($J_z > J_x+J_y$) and for $K=0$ which is the parameter range to which we restrict in the following.  
Furthermore, for the sake of simplicity, we set $J_x=J_y=J$ and $J_z = 1-2J$. 
Along this line, vortex properties of coverings $I$ and $I\!I$ are identical but differ from those of covering  $I\!I\!I$. 
%
%
\subsection{Exact diagonalization}  
\label{ssec_ed}
%
%
For this study, we have also used Majorana fermionization, but contrary to the vortex-free sector, we now have to deal with non-periodic configurations and to perform some finite-size analysis on large systems. Here, we treated system sizes up to $N=16200 \times 16200$.

One important remark is that, using periodic boundary conditions, vortices can only be created in pairs \cite{Kitaev06}. 
Thus, to compute the one-vortex gap, one has first to consider the two-vortex gap  $\Delta E_{\rm 2v}^{(i,j)}$ of a pair of vortices on plaquettes $i$ and $j$ defined by
%
%
\begin{align}
\Delta E_{\rm 2v}^{(i,j)} = \mathcal{E}^{(i,j)} - \gs{0}{}.
\end{align}
%
%
The right-hand side is the energy difference between the ground-state energy $\mathcal{E}^{(i,j)}$ of the considered two-vortex sector and $\gs{0}{}$ which is the ground-state energy of the vortex-free sector. 
Then we use the fact that, in the Abelian phase, vortex excitations are gapped \cite{Kitaev06} so that the interaction of two distant vortices decreases exponentially. 
This allows us to define the one-vortex gap as 
%
%
\begin{align}
\Delta E_{\rm 1v} &=\frac{ \Delta E_{\rm 2v}^{(i,j)}}{2}, \quad {\rm for} \:\:|i-j| \gg 1,
\end{align}
%
%
where it is essential that both vortices are of the same type, \textit{i.~e.}  a or b for model $I\!I$, and a or b or c for model $I\!I\!I$, see Sec.~\ref{sec:themodel}). Practically, results are converged for $|i-j|\simeq 10$ as long as the vortex gap is not too small, namely, away from the transition points (see Sec.~\ref{sec:single_vortices}).

%
%
\subsection{Perturbative approach}  
\label{ssec_pcut}
%
%
To have a better understanding of these few-vortex properties, we have also developped an alternative approach. We computed the spectrum of $H$ perturbatively up to high orders using the perturbative continuous unitary transformation (PCUT) method \cite{Wegner94,Stein97,Mielke98,Knetter00_1,Knetter03_1} around the isolated-dimer limit, where one of the couplings is much larger than the others. Such a strategy has been shown  to be very successful in the original Kitaev model \cite{Schmidt08,Dusuel08_1,Vidal08_2} although restricted to Abelian phases. 
Here, we proceed along the same line as in Ref.~\onlinecite{Vidal08_2} and we skip all technical details which can be found in this reference. 
%
%
\begin{figure}[t]
\includegraphics[width=\columnwidth]{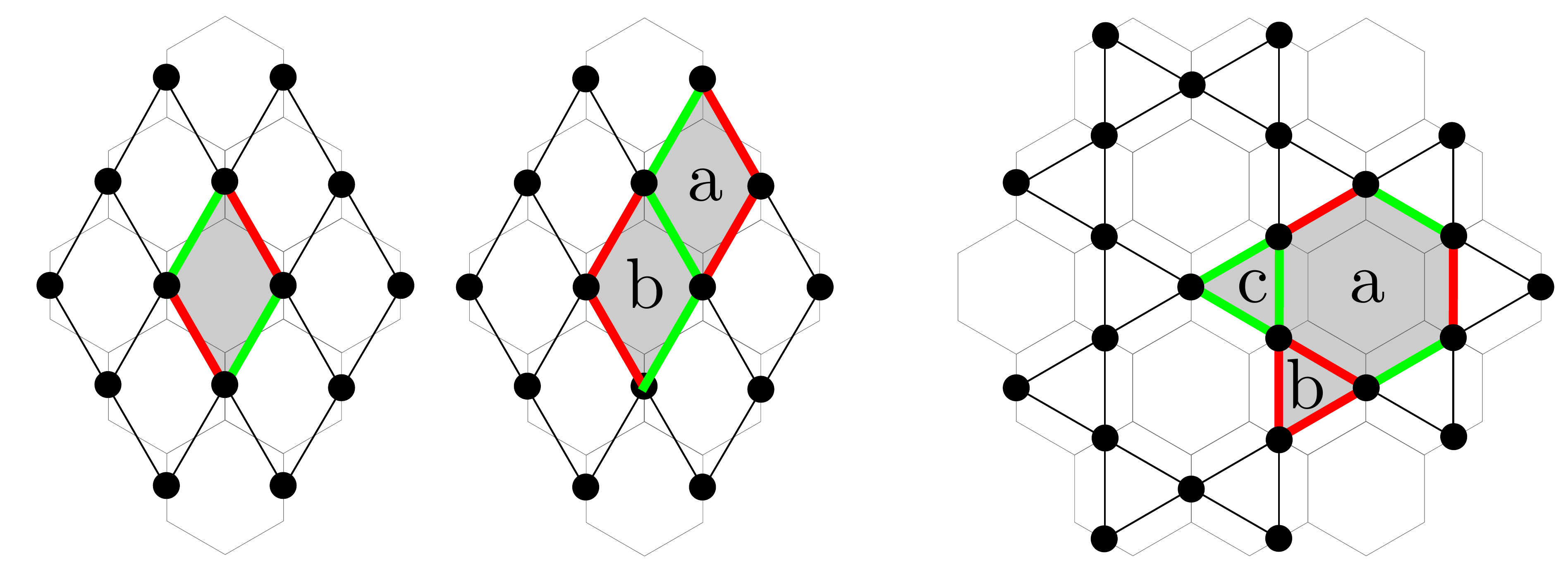}
	\caption{Effective lattices obtained in the isolated-dimer limit $J \ll J_z$ by replacing $z$-dimers by effective sites (filled black circles). The effective lattice for coverings $I$ (left) and $I\!I$ (middle) is a ``square" lattice. For covering $I\!I\!I$ (right), one gets a Kagome lattice. 
Gray-shaded areas show the different types of effective plaquette identified by the same labels as in Fig.~\ref{fig:lattice2}. Red (green) bonds denote $x$ ($y$) links. }
	\label{fig:eff_lattices}
\end{figure}
%
%

As discussed in Refs.~\onlinecite{Kitaev06,Schmidt08,Dusuel08_1,Vidal08_2}, in the limit $J \ll J_z$ it is convenient to 
analyze the problem by replacing $z$-dimers by effective sites with four degrees of freedom corresponding to the four possible dimer states. Then, depending on the dimer configuration, one can face different structures (see Fig.~\ref{fig:eff_lattices}).

For covering $I$,  the effective lattice is a ``square" lattice in which all plaquettes are identical. For covering $I\!I$, one still has a square lattice but one must distinguish between two kinds of plaquettes (a and b) which reflect the operator structure (\ref{eq:Wp_II}).
By contrast, the effective lattice for covering $I\!I\!I$ is a Kagome lattice with three different types (a, b and c) of plaquettes: one hexagonal plaquette with alternating $x$- and $y$-links and two triangular plaquettes with either only $x$- or only $y$-links. It is worth noting that such an effective lattice was also obtained in Ref.~\onlinecite{Dusuel08_2}, starting from a different initial geometry. 
  
As already explained, in the Abelian phase the Kitaev honeycomb model contains two different kinds of excitations: low-energy vortices and high-energy fermions. 
The PCUT method around the  isolated-dimer limit provides (order by order in $J/J_{\rm z}$ and in the thermodynamic limit) an effective Hamiltonian which commutes with the number of fermions. Thus, the low-energy spectrum is obtained by considering the zero-fermion (quasiparticle) block  usually denoted 0QP. In this model, the effective Hamiltonian in the 0QP sector is particularly simple because it can be fully expressed in terms of the conserved $\mathbb{Z}_2$ plaquette operators ${W}_p$. 
Its general structure reads
%
%
\begin{equation}
 H_{\rm eff}^{\mathrm{0QP}} = E_0 - \sum_{\{p_1,..,p_n\}} C_{p_1,\ldots , p_n}{W}_{p_1} \ldots{W}_{p_n} , 
\end{equation}
%
%
where $E_0$ is a constant and where $\{p_1,p_2,\ldots,p_n\}$ denotes a set of $n$ plaquettes. The sum runs over all possible plaquette numbers and configurations. In the present work, we have computed the corresponding coefficients $C_{p_1,\ldots , p_n}$ of this multi-plaquette expansion up to order 10. The ground-state energy of any vortex configuration is then readily obtained  by replacing $W_p$ by $-1$ if there is a vortex and by $+1$ otherwise.

Before presenting the results, we would like to point out that the validity range of the PCUT results does \emph{not} depend on the vortex filling. However, this perturbative treatment is only valid as long as all fermionic gaps are finite. 
For instance, this implies that although the gapless phase is reduced to a single point for dimer covering $I\!I\!I$, this perturbative approach breaks down well before this point is reached since some fermionic gaps in other vortex sectors vanish \cite{Kamfor10}. 
For the three dimer coverings considered here, we are led to conjecture that there always exists a vortex configuration such that the fermionic gap vanishes for $J_z=J_x+J_y$. For covering  $I$ and $I\!I$, this is achieved in the vortex-free sector but, for covering $I\!I\!I$, this is found for more complex vortex configurations \cite{Kamfor10}. As a consequence, we restrict our analysis to $J\in [0,1/4]$.

%
%
\subsection{Single-vortex properties}
\label{sec:single_vortices}
%
%
%
On the line considered here ($J_x=J_y=J$), coverings $I$ and $I\!I$ are identical. However, for covering 
$I\!I\!I$, one must distinguish between two different vortex gaps according to the kind of plaquette which is excited  (triangle or hexagon of the effective Kagome lattice). 
%
%
\begin{figure}[t]
        \includegraphics[width=0.8\columnwidth]{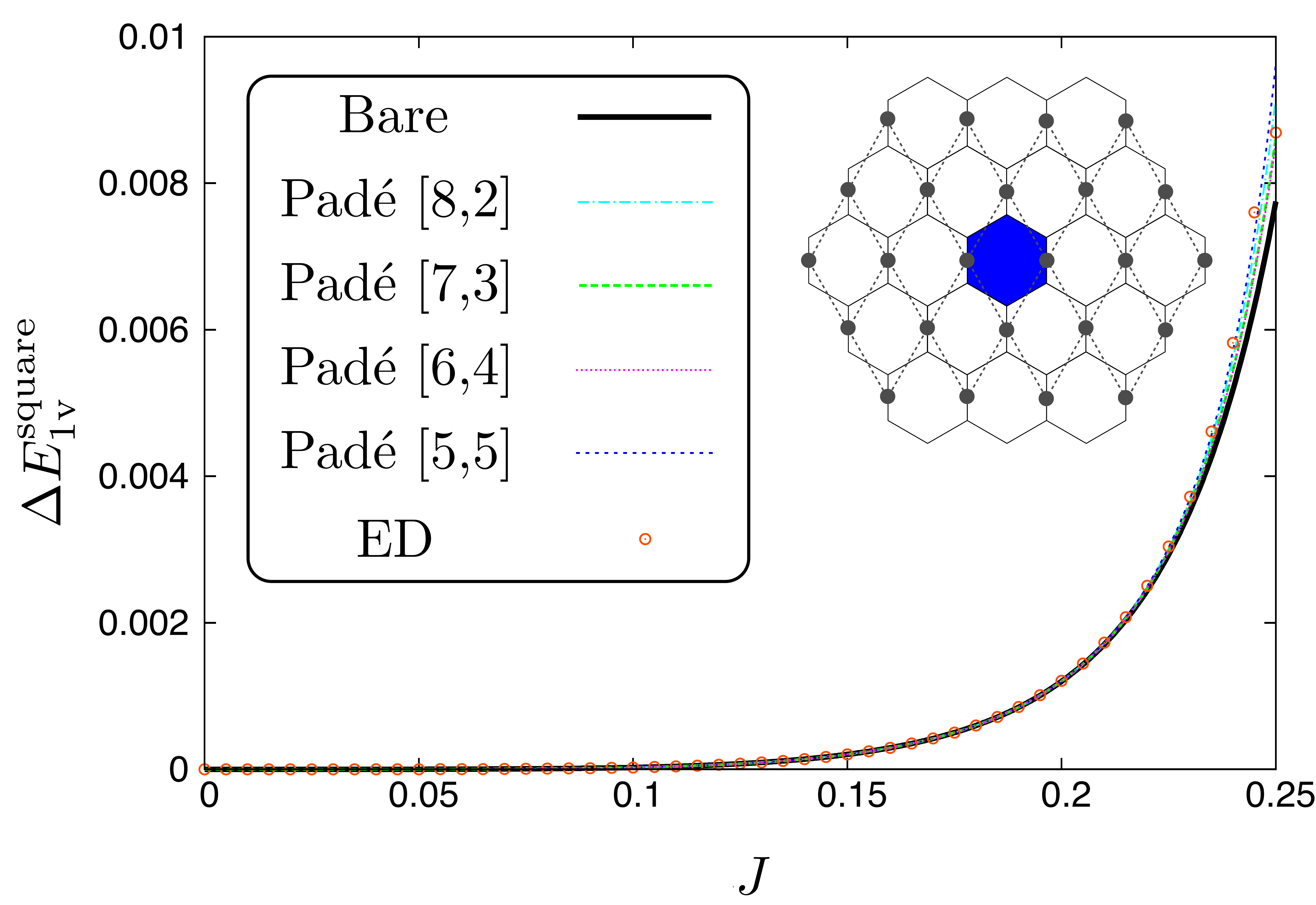}
	\includegraphics[width=0.8\columnwidth]{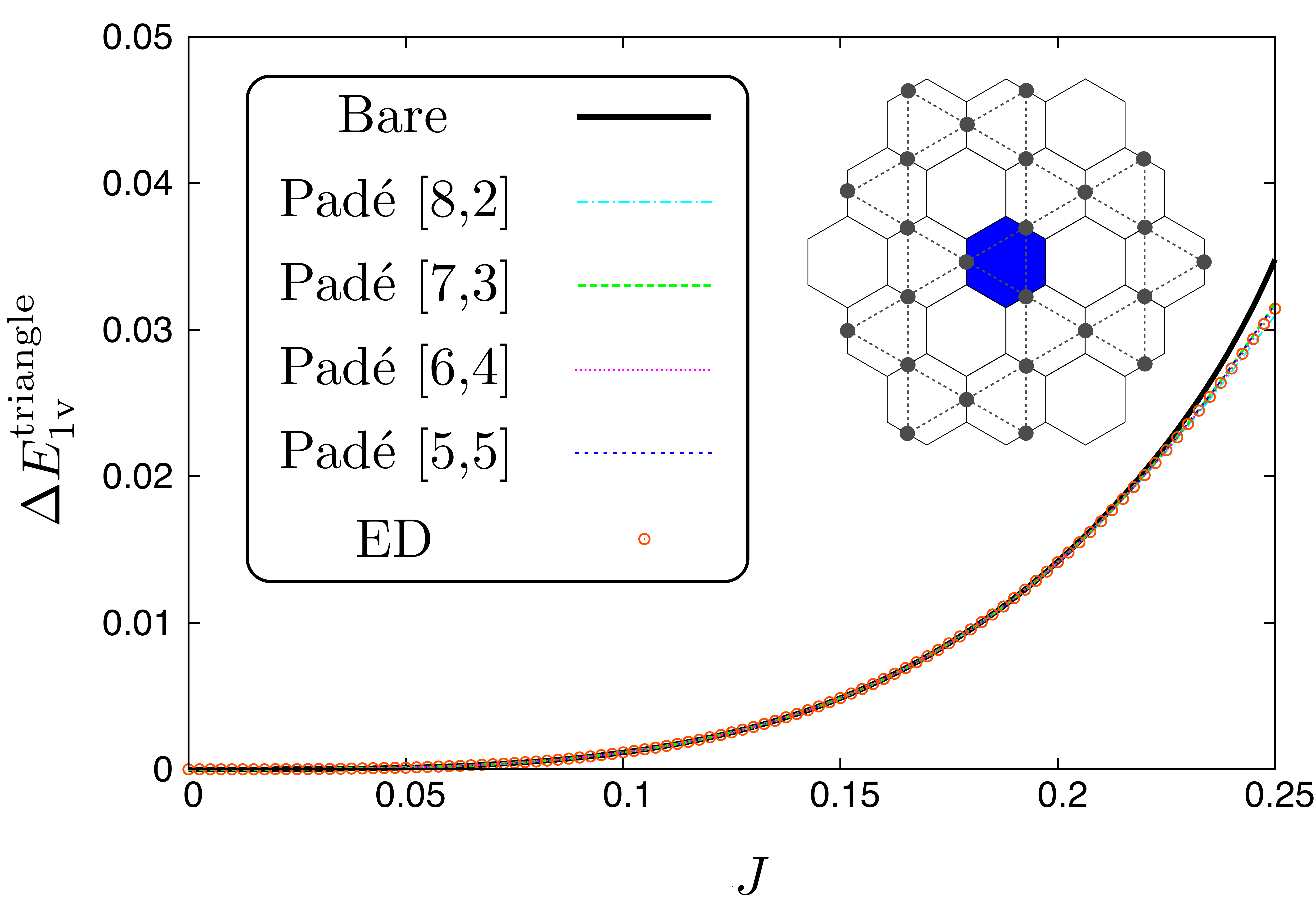}
	\includegraphics[width=0.8\columnwidth]{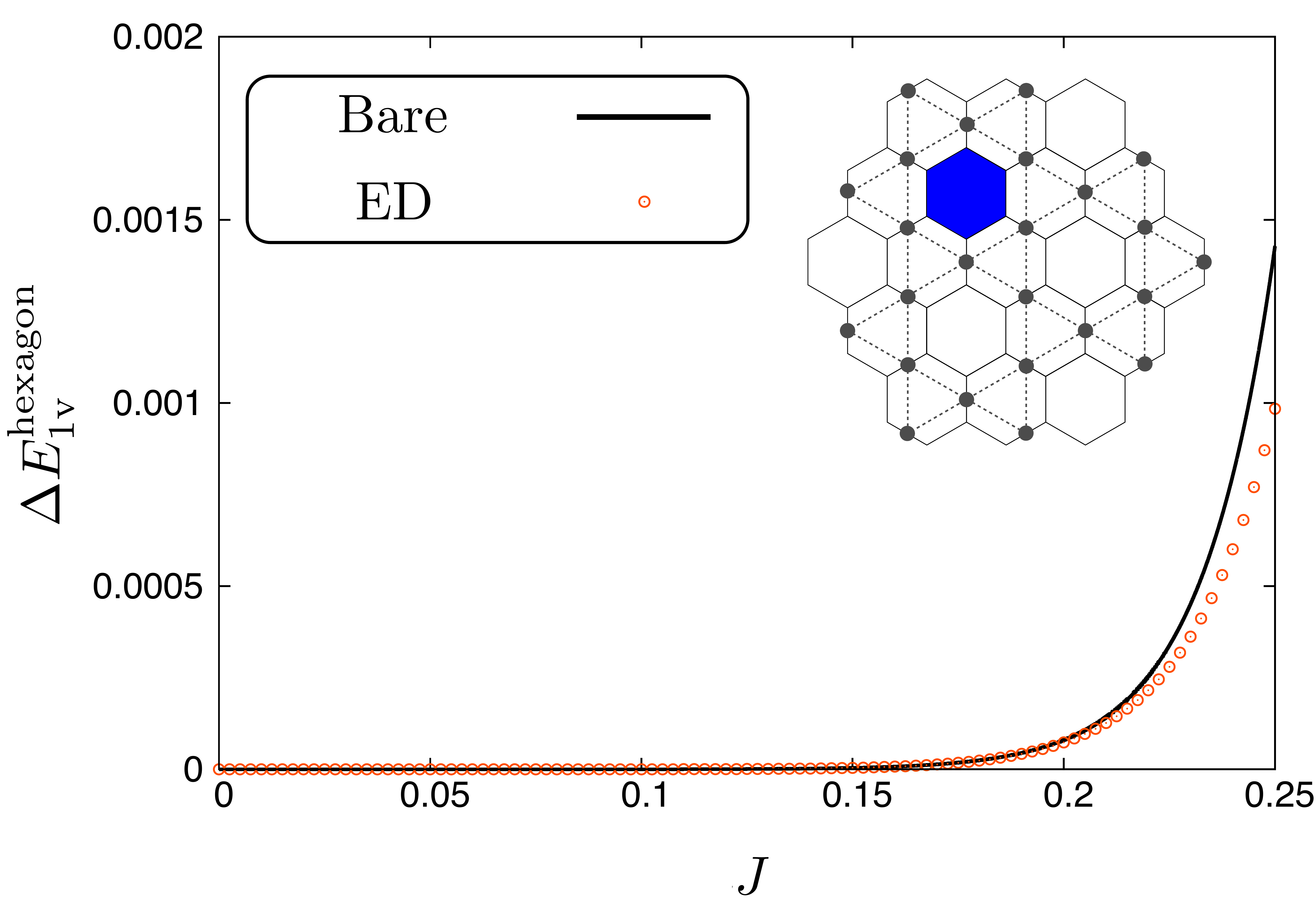}
	\caption{Single-vortex gap $\Delta E_{\rm 1v}$ as a function of $J$ for coverings $I$ and $I\!I$ (top) and for covering $I\!I\!I$ (middle and bottom). We give the bare series expansion computed up to order 10, some Pad\'e approximants of these series (except for the hexagonal plaquette for which we do not get enough terms in the series to provide reliable approximants), and the exact diagonalization results.  Insets show the excited plaquette considered 
together with the effective lattice.}
	\label{fig:single_vex_full}
\end{figure}
%
%
The single-vortex gaps obtained by Majorana fermionization and by the PCUT are displayed for these three different cases in Fig.~\ref{fig:single_vex_full}.  An excellent agreement between both approaches is evidenced. 

The most striking result is that the value of this vortex gap depends strongly on the type of effective plaquette. This can be understood easily by noting that the leading order in the perturbative expansion is determined by the ``perimeter'' of the plaquette (3 for the triangle, 4 for the square, and 6 for the hexagon). In this sense, one may say that the mass of the vortex is strongly sensitive to the dimerization pattern.

For illustration, we give the analytical expression of these various gaps computed up to order 10
\begin{eqnarray}
\label{eq:vortex_gap_square}
\Delta E_{\rm 1v}^{\rm square} &=& \frac{1}{8}J^4+\frac{3}{4}J^5+\frac{13}{4}J^6 + \frac{25}{2}J^7 +\frac{5835}{128}J^8 \nonumber \\
&&+\frac{10381 J^9}{64} +\frac{18277}{32}J^{10},
\end{eqnarray}
\begin{eqnarray}
\label{eq:vortex_gap_triangle}
\Delta E_{\rm 1v}^{\rm triangle} &=& \frac{3}{4}J^3 + 3 J^4+\frac{531}{64}J^5 + \frac{273}{16}J^6 + \frac{2379}{128}J^7 \nonumber \\
&& -\frac{47961}{1024}J^8 -\frac{6570213}{16384}J^9 -\frac{55469271}{32768}J^{10},\nonumber \\
\end{eqnarray}
\begin{eqnarray}
\label{eq:vortex_gap_hexagon}
\Delta E_{\rm 1v}^{\rm hexagon} &=& \frac{3}{128}J^6 + \frac{81}{256}J^7 + \frac{2721}{1024}J^8 + \frac{36249}{2048}J^9 \nonumber \\
&& + \frac{1652919}{16384}J^{10} .
\end{eqnarray}

Finally, let us remind that in this Abelian phase, vortices behave as Abelian anyons or hard-core bosons depending on their relative position (see Ref.~\onlinecite{Vidal08_2} for a detailed discussion). In the present case, for covering $I\!I$, plaquette excitations (a) and (b) are mutual semions whereas they behave, individually, as hard-core bosons. 
For covering $I\!I\!I$, triangular and hexagonal plaquette excitations are mutual semions whereas triangular (hexagonal) plaquette  excitations behave individually as hard-core bosons.
%
%
\subsection{Vortex-vortex interactions}
\label{sec:two_vortices}
%
%
%
To conclude this section, let us discuss the nature of vortex-vortex interactions $v^{(i,j)}$ between two vortices on plaquettes $i$ and $j$. The interaction energy between two such vortices is defined as 
%
%
\begin{align}
v^{(i,j)}  = \Delta E_{\rm 2v}^{(i,j)}-\Delta E_{\rm 1v}^{(i)} -\Delta E_{\rm 1v}^{(j)},
\end{align}
%
%
where the energy of each type of vortex is subtracted.

\begin{figure}[h]
	\centering
	\includegraphics[width=0.4\columnwidth]{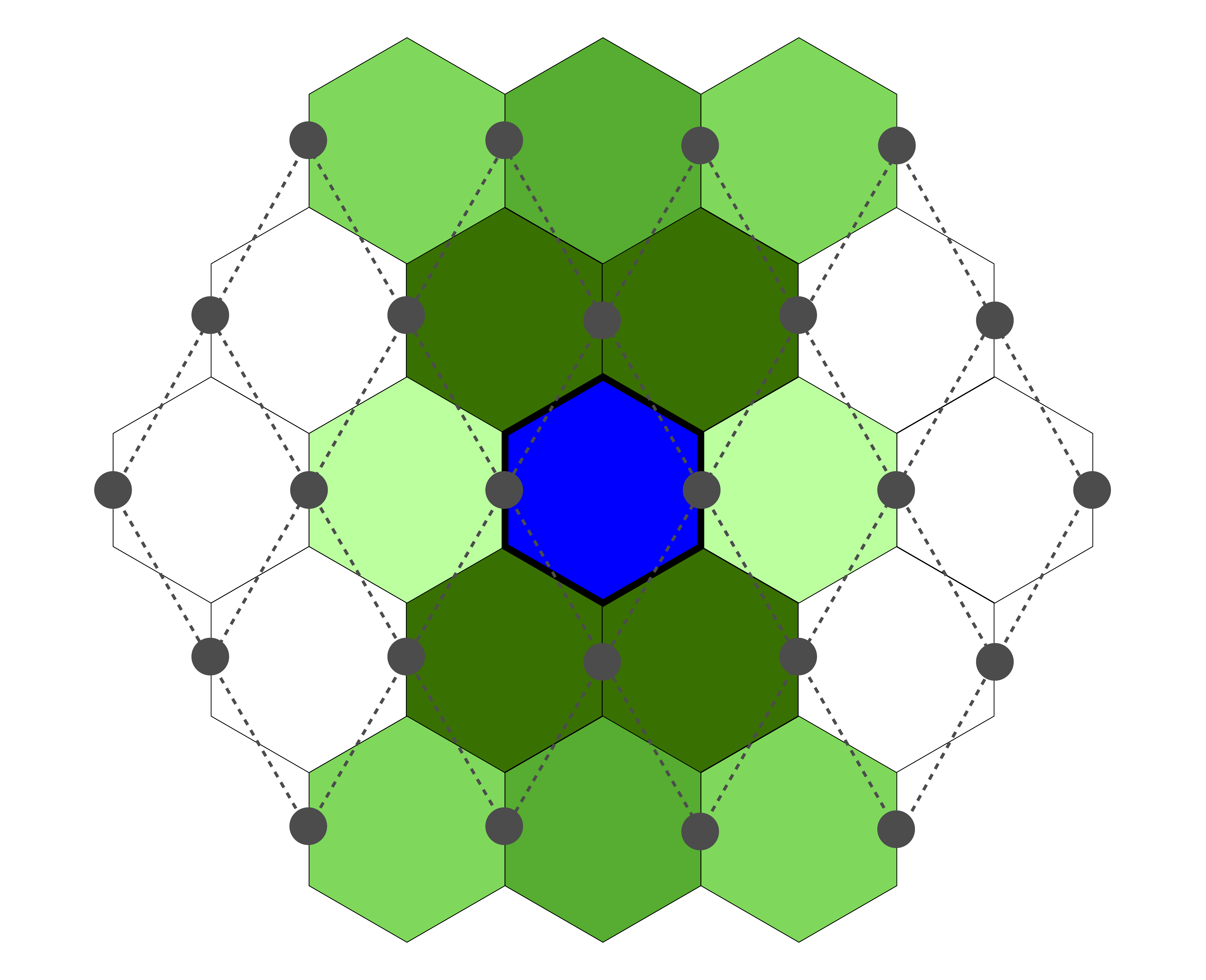}
	\includegraphics[width=0.4\columnwidth]{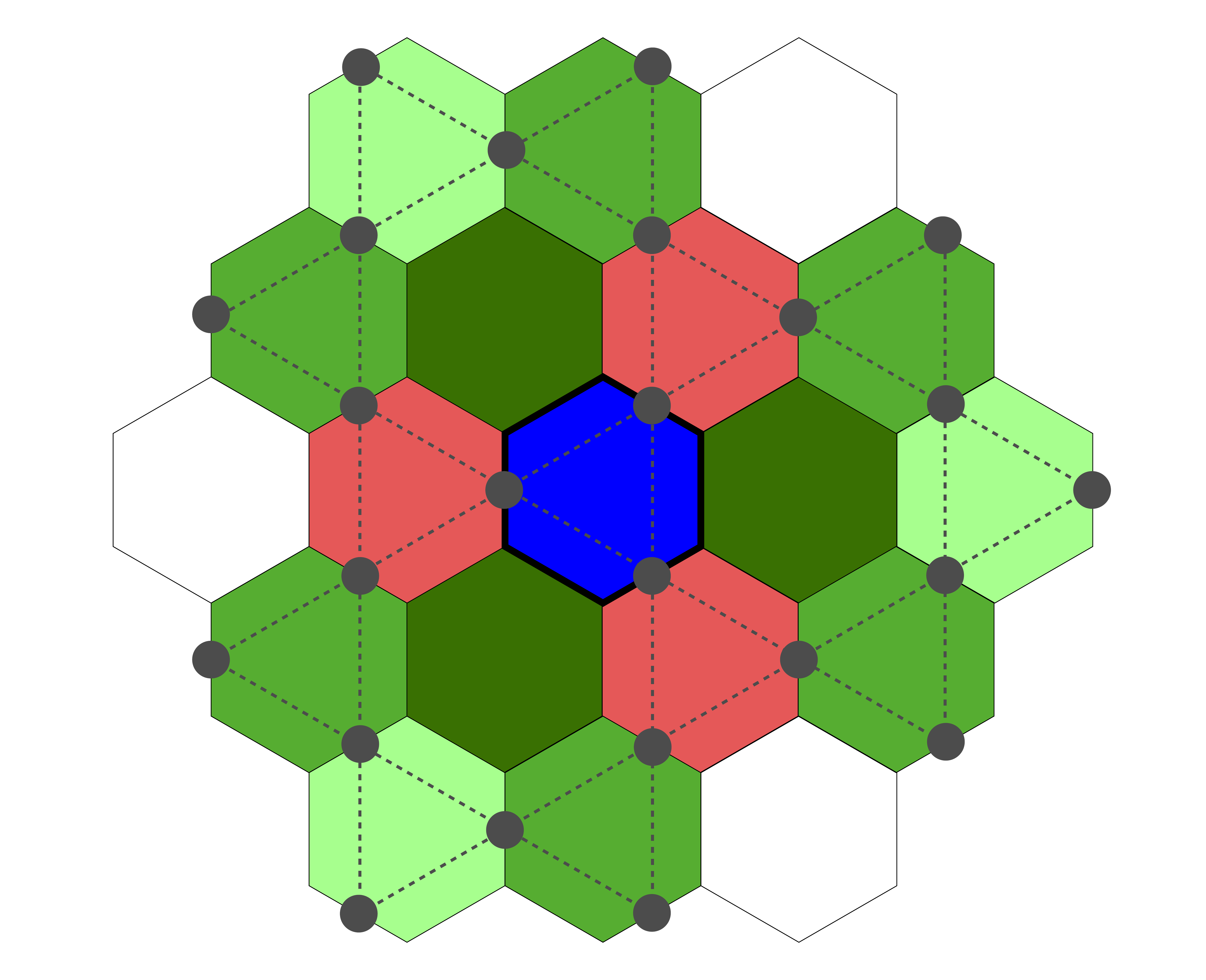}
	\caption{Schematic view of vortex-vortex interactions $v^{(i,j)}$ for coverings $I$ and $I\!I$ (left) and for covering $I\!I\!I$ (right). A green (red) plaquette interacts attractively (repulsively) with a vortex placed on the central blue plaquette. The color code is such that a darker green corresponds to a stronger attractive interaction.}
	\label{fig:vortex_interactions}
\end{figure}

Of course, such an interaction depends on the relative position of the vortices. Once again, on the line considered here ($J_x=J_y$) this interaction is the same for coverings $I$ and $I\!I$. For these two coverings, we found that this interaction is always attractive (this is also true for $J_x \neq J_y$) whereas, for covering $I\!I\!I$, there are some positions for which it is repulsive. These results are illustrated qualitatively in Fig.~\ref{fig:vortex_interactions}. 

%
%
%
%
%

\begin{figure}[t]
	\centering
        \includegraphics[width=0.8\columnwidth]{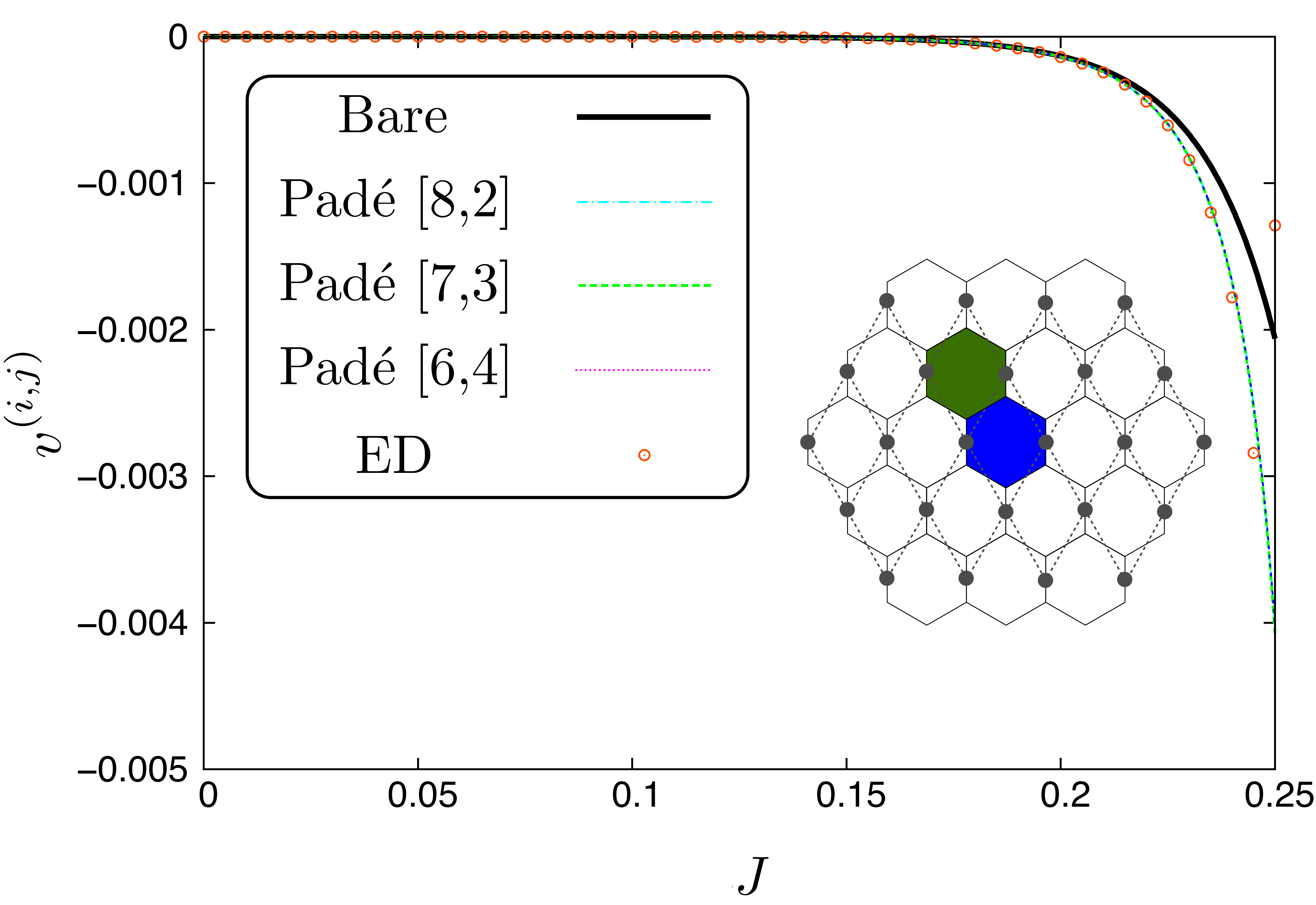}
	\includegraphics[width=0.8\columnwidth]{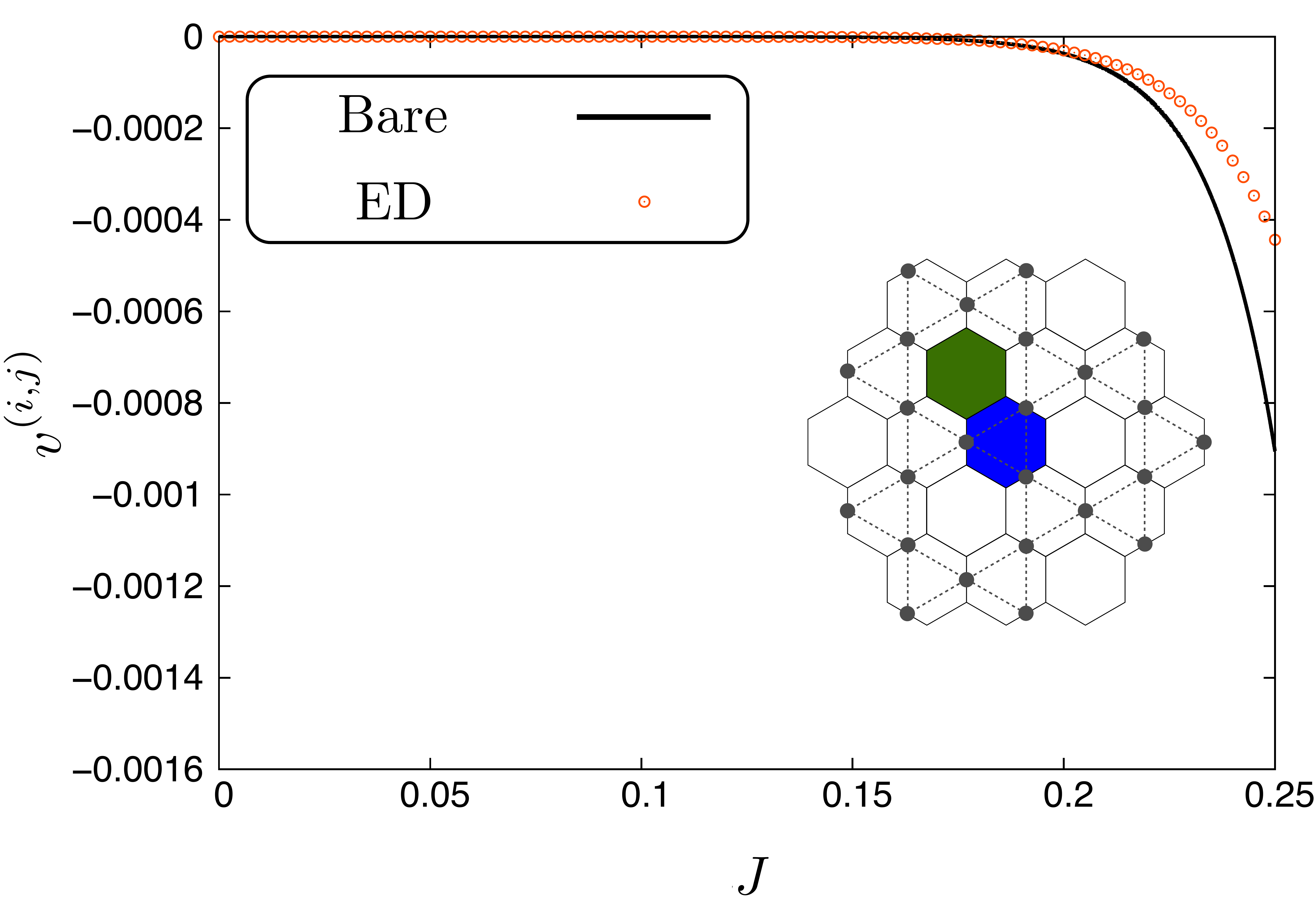}
	\includegraphics[width=0.8\columnwidth]{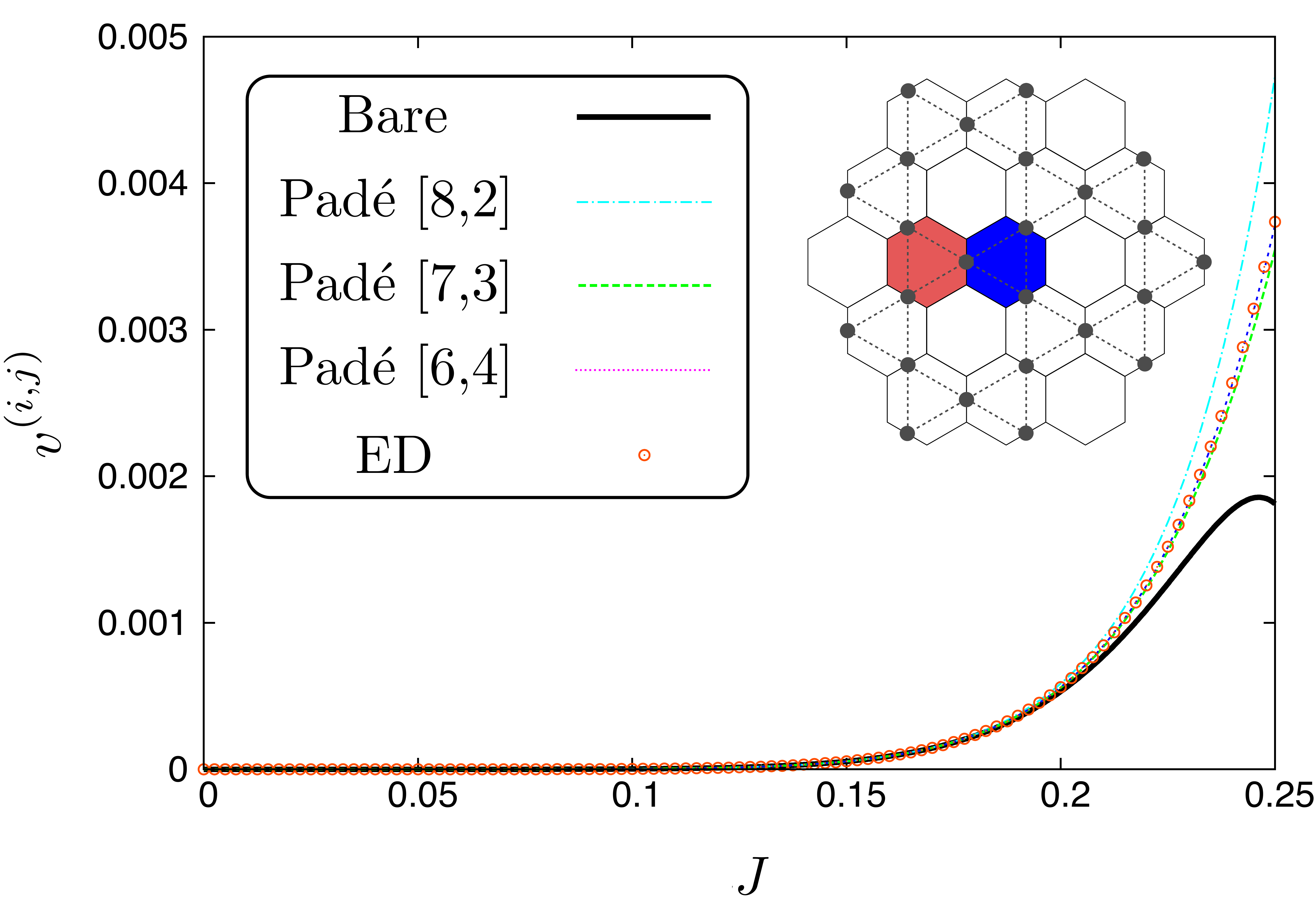}
	\caption{Vortex-vortex interaction $v^{(i,j)}$ for  two adjacent effective plaquettes as a function of $J$ for  coverings $I$ and $I\!I$ (top) and for covering $I\!I\!I$ (middle and bottom). 
 As in Fig.~\ref{fig:single_vex_full}, we do not show Pad\'e approximants when hexagonal plaquettes are involved since the corresponding series starts at order 6 so that there is not enough terms to get reliable results when resumming.
 Insets show the nearest-neighbor excited plaquettes for which the two-vortex interaction is computed.}
 \label{fig:ia_full}
\end{figure}

In addition, Fig.~\ref{fig:ia_full} shows a comparison between exact diagonalizations and PCUT results for some special cases where vortices are placed on nearest-neighbor effective plaquettes.  
Again, both approaches are in very good agreement and confirm that some repulsion can be found in covering $I\!I\!I$.
This interesting finding can be  readily extracted from the series expansion obtained from PCUT. Indeed, the leading order of the interaction between two effective triangular plaquettes is given by the repulsive term $+7 J^6/8$ while the dominant contribution to the interaction between a triangular and a hexagonal plaquette is attractive and reads $- 7J^7/256$. 

%
%
%
%

%
\section{Summary and perspectives}
\label{sec:summary}
%
%
We have extended the Kitaev honeycomb model to arbitrary dimer coverings satisfying matching rules. This results in a family of models which can still be solved exactly by Majorana fermionization. Although these models are equivalent at the isotropic point, they strongly differ at any other point in parameter space. 

In the present work, we focused on three different coverings ($I$, $I\!I$, and $I\!I\!I$) having the smallest possible unit cells.  For these three models, we have studied the ground-state phase diagram as a function of the three-spin interaction strength $K$. We found that there are many similarities between the phase diagrams of $\h$ and $\hh$. However, the phase diagram of $\hhh$ is dramatically different. In particular, the gapless phase at $K=0$ is reduced to a single point and, for large $K$, no Abelian phase exists. 
We have furthermore studied one- and two-vortex properties in these different  coverings in the Abelian phase at $K=0$ and we have shown that they also depend on the covering considered. These findings have been analytically confirmed by perturbative expansions around the isolated-dimer limit inside the Abelian phase.

Finally, let us point out two routes left for the future. Firstly, it would be interesting to investigate other vortex conﬁgurations since, as recently observed in the original Kitaev model \cite{Lahtinen10}, vortex interactions could lead to new topological phases characterized, notably, by exotic Chern numbers \cite{Kitaev06,Gils09,Ludwig10,Kamfor10}. Secondly, it would be worth studying the effect of disordered dimer and/or vortex configurations to see if an Anderson-type transition can occur in such systems. Note that other kinds of disorder have recently been considered in Refs.~\onlinecite{Willans10,Dhochak10}. We hope that the present work will stimulate further investigations in a near future.

\acknowledgments KPS acknowledges ESF and EuroHorcs for funding through his EURYI.     


\end{document}